\documentclass[preprint]{elsarticle}

\usepackage{lineno}
\modulolinenumbers[5]

\usepackage[english]{babel}
\usepackage[utf8]{inputenc}
\usepackage[T1]{fontenc}

\usepackage[a4paper,top=3cm,bottom=2cm,left=3cm,right=3cm,marginparwidth=1.75cm]{geometry}

\usepackage{amsmath}
\usepackage{textcomp, gensymb}
\usepackage{natbib}
\usepackage{graphicx}
\usepackage[colorinlistoftodos]{todonotes}
\usepackage[colorlinks=true, bookmarks=false, allcolors=blue]{hyperref}
\usepackage{float}
\usepackage{url}
\usepackage{diagbox}
\usepackage{multirow}
\usepackage{exscale}
\usepackage{relsize}
\usepackage{verbatim}

\usepackage{pdflscape}
\usepackage{lscape}
\usepackage{lineno}

\usepackage{array}
\usepackage{booktabs}
\usepackage{float}
\usepackage{ragged2e}
\usepackage{supertabular}
\usepackage[titletoc]{appendix}
\usepackage[tableposition=top]{caption}
\usepackage[justification=justified,singlelinecheck=off]{caption}

\makeatletter
\newcommand\fake@math{}
\def\fake@math#1\){[math]}
\makeatother

\newcommand{\beginsupplement}{%
        \setcounter{table}{0}
        \renewcommand{\thetable}{S\arabic{table}}%
        \setcounter{figure}{0}
        \renewcommand{\thefigure}{S\arabic{figure}}%
     }

\biboptions{authoryear}

\makeatletter
\def\ps@pprintTitle{%
   \let\@oddhead\@empty
   \let\@evenhead\@empty
   \let\@oddfoot\@empty
   \let\@evenfoot\@oddfoot
}
\makeatother

\begin{document}
\title{Prospects for Mapping Temporal Height Variations of the Seasonal CO$_2$ Snow/Ice Caps at the Martian Poles by Co-registration of MOLA Profiles}

\captionsetup[figure]{labelfont={bf},name={Fig.},labelsep=period}
\captionsetup[table]{labelfont={bf},labelsep=period}


\renewcommand{\vec}[1]{\boldsymbol{#1}}
\newcommand{\veci}[2]{\boldsymbol{#1}_{\mathrm{\small #2}}}
\newcommand{\veciu}[2]{\boldsymbol{\mathrm{#1}}^{\mathrm{\small #2}}}
\newcommand{\scli}[2]{#1_{\mathrm{\small #2}}}
\newcommand{\tof}{\tau}
\newcommand{\sclii}[2]{\mathrm{#1}_{\mathrm{#2}}}

\begin{frontmatter}

\author[1]{Haifeng Xiao\corref{cor1}}
\ead{Haifeng.Xiao@campus.tu-berlin.de}
\author[2]{Alexander Stark}
\author[3]{Gregor Steinbr\"ugge}
\author[4]{Robin Thor}
\author[5,6]{Frédéric Schmidt}
\author[1,2]{Jürgen Oberst}

\cortext[cor1]{Corresponding author}

\address[1]{Institute of Geodesy and Geoinformation Science, Technische Universität Berlin, Berlin, Germany}
\address[2]{Institute of Planetary Research, German Aerospace Center (DLR), Berlin, Germany}
\address[3]{Department of Geophysics, Stanford University, Stanford, USA}
\address[4]{Institute of Atmospheric Physics, German Aerospace Center (DLR), Oberpfaffenhofen, Germany}
\address[5]{Université Paris-Saclay, CNRS, GEOPS, Orsay, France}
\address[6]{Institut Universitaire de France (IUF), Paris, France}

\begin{abstract}
We investigate the feasibility and demonstrate the merits of using Mars Orbiter Laser Altimeter (MOLA) profiles to retrieve seasonal height variations of CO$_2$ snow/ice cap in Mars' polar areas by applying a co-registration strategy. 
We present a prototype analysis on the research region of [85.75$\degree$S, 86.25$\degree$S, 300$\degree$E, 330$\degree$E] that is located on the residual south polar cap. 
Our method comprises the recomputation of MOLA footprint coordinates with an updated Mars Global Surveyor (MGS) ephemeris and a revised Mars rotation model. 
The reprocessed MOLA dataset at the South Pole of Mars (poleward of 78$\degree$S) is then self-registered to form a coherent reference digital terrain model (DTM). 
We co-register segments of reprocessed MOLA profiles to the self-registered MOLA reference DTM to obtain the temporal height differences at either footprints or cross-overs. 
Subsequently, a two-step Regional Pseudo Cross-over Adjustment (RPCA) procedure is proposed and applied to post-correct the aforementioned temporal height differences for a temporal systematic bias and other residual errors. 
These pseudo cross-overs are formed by profile pairs that do not necessarily intersect, but are connected through the underlaying DTM.
Finally, CO$_2$ snow/ice temporal height variation is obtained by median-filtering those post-corrected temporal height differences. 
The precision of the derived height change time series is $\sim$4.9~cm. 
The peak-to-peak height variation is estimated to be $\sim$2~m. 
In addition, a pronounced "pit" (transient height accumulation) of $\sim$0.5~m in magnitude centered at $\sclii{L}{s}=210\degree$ in southern spring is observed. 
The proposed method opens the possibility to map the seasonal CO$_2$ snow/ice height variations at the entire North and South polar regions of Mars.
\end{abstract}

\begin{keyword}
Mars \sep CO$_2$ ice \sep height variation \sep MOLA \sep co-registration \sep pseudo cross-over adjustment
\end{keyword}

\end{frontmatter}

\newpage

\section{Introduction}
\label{sec:intro}

\quad Every Martian year, up to one third of the atmosphere's CO$_2$ mass is involved in interactions with the polar surface through a seasonal condensation/sublimation process \citep{Leighton66}. 
In winter, when temperature drops below the CO$_2$ condensation point, the CO$_2$ solidifies and accumulates as snow or frost in the polar regions of Mars.
During this time, the seasonal CO$_2$ polar caps can extend down to 50$\degree$S/N \citep{piqueux15}. 
In spring, the CO$_2$ sublimates back into the atmosphere \citep{smith2001co2, aharonson2004depth, matsuo2009density, Schmidt2009}. 
Thus, during a Martian year, CO$_2$ migrates from North Pole to South Pole and vice versa. 
Accurate measurements of seasonal Martian polar CO$_2$ snow/ice height and volume variation can serve as crucial constraints on the Mars climate system. 
Such measurements can also help to prepare for future robotic and human exploration missions to the planet. 
The mass of the seasonal polar caps condensed onto the Martian surface has been measured by gravity variation \citep{smith09, genova16} and neutron and gamma ray flux \citep{feldman03, prettyman09}, and modeled by General Circulation Models \citep[GCMs, e.g.,][]{smith99} and energy balance model \citep[e.g.,][]{kieffer01, Schmidt2010}. 
Meanwhile, the direct height variation measurements of the seasonal polar caps have been made by the Mars Orbiter Laser Altimeter (MOLA) \citep{smith2001co2, aharonson2004depth, Jian2009, smith18}, rock shadowing \citep{mount15}, and Bayesian inversion involving a radiative transfer model and imaging spectroscopy \citep{andrieu18}. 
While the latter two are spatially and temporally limited due to data availability, MOLA estimates can be easily extended to cover the entire polar regions. 
The first quantitative measurement of the local seasonal CO$_2$ snow/ice height variation was made by MOLA. 
\cite{smith2001co2} measured the peak-to-peak seasonal height variations to be $\sim$1$\pm$0.14~m at both of the poles based on profile analysis at annuli, averaged over longitude. 
\cite{aharonson2004depth} used temporal height differences at the intersections of MOLA ground-tracks, i.e., cross-overs, and fitted sinusoidal model functions including annual and semi-annual terms to them at grid bins of size 3$\degree$ in longitude and 0.5$\degree$ in latitude. 
The authors showed that the maximum seasonal snow/ice height variations can reach greater (and different compared to \cite{smith2001co2}) levels of $\sim$1.5$\pm$0.21~m at the North Pole and $\sim$2.5$\pm$0.21~m at the South Pole. 
Unfortunately, the results are noisy and time-resolved temporal height variations at local scales remain unexplored. 
Moreover, a reliable long-term time series of surface height change is needed to confirm or reject the reported retreat and thinning of the perennial CO$_2$ ice layer of the residual south polar cap which is up to~10~m in thickness \citep{malin2001Observational, blackburn2010Sublimation, thomas2016mass}.

\quad The primary goal of MOLA was the global mapping of the Martian topography for the purpose of geological, geophysical, and climate modeling studies \citep{zuber1992mola}. 
The experiment was not designed to map temporal height changes in the topography, unlike e.g., the Ice, Cloud and Land Elevation Satellite (ICESat) Geoscience Laser Altimeter System (GLAS) or CryoSat-2 for the cryosphere of Earth \citep{schroder2019}. 
On Earth, the joint inversion for topography and height change parameters within a moving box region along repeat tracks is a commonly used technique \citep{pritchard2009, schenk2012, schroder2019}. 
However, the MOLA orbit does not feature repeat tracks that are specifically devoted to height change mapping. 
\cite{smith2001co2} and \cite{aharonson2004depth} approached the problem of detecting height changes by using temporal differential measurements at intersections of the ground tracks. 
However, this method can suffer from significant errors due to interpolation between footprints forming the cross-over, when the distance between consecutive footprints is large ($\sim$300~m for MOLA footprints $\sim$170~m in diameter) and where the terrain is rough. 
Furthermore, for planetary laser altimeters, the geolocation accuracy of the acquired profiles is often compromised by significant unmodeled orbit and attitude errors, clock drift and aging errors, laser boresight alignment errors, and planetary rotational modelling errors. 
These errors can translate into lateral shifts of the laser profiles and further contaminate the height differences at cross-overs \citep[see also][]{steinbrugge2015tide, steinbrugge2018bela}. 
Besides, this approach cannot be applied generally in Solar System Exploration, since grazing angles between intersecting ground tracks for laser altimeter on a polar orbit around slowly-rotating bodies as it is the case, e.g., for the Mercury Laser Altimeter (MLA), pose an unfavorable geometry for the estimation of cross-over height differences \citep{mazarico2010cross-over}. 
Finally, ground tracks merge at the poles for laser altimeters in polar orbits (e.g., MOLA and Lunar Orbiter Laser Altimeter (LOLA)), causing a greatly limited number of cross-overs towards the equator where the tracks run almost parallel to each other.  

\quad Here we propose the co-registration between local segments of laser profiles and Digital Terrain Models (DTMs) from self-registered laser profiles as a new approach for the detection of surface height change. 
Co-registration is a post-processing technique for aligning the laser profiles to a continuous DTM. 
This procedure is not limited to cross-overs and can use both, information from all footprints along the local segments of profiles, and the topographic information from the DTMs, which are used as reference.
In \ref{sec:co-regist_self_regist}, we present this co-registration technique with different parameterizations and introduce the concept of self-registration of laser profiles to a reference DTM. 
In this paper, we introduce the local co-registration strategy 
and showcase its merits by applying it to resolving the seasonal height change of the CO$_2$ snow/ice deposits. 

\quad In their studies of seasonal polar snow/ice height variations, \citet{smith2001co2} found a global systematic temporal bias in MOLA footprint heights which features an amplitude of 50~cm and a phase that matches the synodic period of Mars. 
In the final MOLA data product this bias was removed by means of a cross-over adjustment \citep{neumann2001cross-over}.
To correct for this artifact in their height variation measurement, \citet{smith2001co2} subtracted the height change time series at 60$\degree$S/N (where limited height variation is expected) from the time series obtained at the polar regions. 
Other effects, such as surface deformation due to snow/ice load, tidal deformation, and dark material erupted of the cold jets, are at <1~cm and can thus be left unattended \citep{kieffer2006jetting, matsuo2009density, turyshev2010laserranging}. 
To correct for this systematic temporal artifact and residual errors in the reprocessed MOLA profiles and self-registered MOLA reference DTMs that are involved in the aforementioned local co-registration processes, a novel post-correction procedure is proposed in this study. 
It makes use of DTM-based "pseudo cross-overs" \citep{barker2016lola} acquired within 5 days in a process as regional pseudo cross-over adjustment (RPCA). 
As the height misfit at a pseudo cross-over is assigned as the difference in height corrections in the local co-registrations of the profile segment pair, that form the pseudo cross-over, to the underlying self-registered MOLA reference DTM, the aforementioned disadvantages of using height differences at cross-overs can be avoided. 
In addition, since the profile segment pair forming a pseudo cross-over do not necessarily need to intersect, the available number of pseudo cross-overs can be boosted by several times and thus offer much more constraints in the adjustment. 
Meanwhile, as the profile segment pairs forming the pseudo cross-overs can be widely distributed across the research region, the offered constraints are also more global, especially if the research region is extensive.

\quad The rest of the paper is structured as follows: In Sec.~\ref{sec:data}, we introduce the MOLA Precision Experiment Data Records (PEDR) dataset. 
It is followed by the reprocessing of the MOLA PEDR datset and the self-registration of these reprocessed MOLA profiles to obtain a coherent reference DTM at the South Pole (Sec.~\ref{sec:mola_rep}). 
In Sec.~\ref{sec:methods}, the proposed local co-registration strategy to generate temporal height differences at either footprints or cross-overs, and the post-correction procedure are detailed. 
Subsequently, the results from these methods either with or without the local co-registration strategy are presented and compared in Sec.~\ref{sec:res}. 
Other factors that can affect the performance of the proposed methods, and their potential applications are discussed in Sec.~\ref{sec:disc}. 
Finally, conclusions are drawn in Sec.~\ref{sec:conc}.

\section{MOLA PEDR dataset} 
\label{sec:data}

\quad Launched on November 1996 onboard the Mars Global Surveyor (MGS), MOLA was the first laser altimeter outside of the Earth-Moon system. 
MOLA measured the topography of Mars within approximately footprints of 160~m in diameter with a center-to-center along-track footprint spacing of around 300~m in the nadir configuration \citep{zuber1992mola}. 
MOLA ceased altimetry operations on June 30, 2001 after a clock oscillator failure but continued to operate for passive radiometry measurements through the end of the MGS mission in 2006. 
MOLA laser profiles have been extensively used in geological and geophysical applications for the past two decades \citep[e.g.,][]{kreslavsky2000roughness, heavens2017mola, parro2017heat}.

\quad The MOLA PEDR dataset Version L \citep{pds-geosciences.wustl.eduMOLA} includes laser profiles acquired in the mapping phase and the extended phases lasting from March 1999 to June 2001 which spans slightly more than a Martian year (from MY24 to MY25). 
This dataset contains shot emission times, one-way ranges, quality flags, range corrections on the order of meters due to the electronics delays and range walk, as well as other instrument and observation characteristics. 
\cite{neumann2001cross-over} has carried out a global cross-over analysis with corrections parameterized by slowly-varying functions to adjust the tracks in along-track, cross-track, and radial dimensions to post-correct for residual orbit, timing, and pointing errors. 
These corrections in longitude, latitude, and height for each footprint are also provided in the MOLA PEDR dataset. 
The radial accuracy of individual footprints after the global cross-over adjustment is approximately 1~m, while lateral geolocation errors are estimated to be within 100~m in along-track and cross-track directions \citep{neumann2001cross-over}. 
The MOLA Mission Experiment Gridded Data Records (MEGDRs) are gridded global topographic maps of Mars created by binning altimetry values from the MOLA PEDR dataset \citep{smith2003megdr}.

\section{MOLA reprocessing and self-registration}
\label{sec:mola_rep}

\quad We first reprocessed the MOLA PEDR dataset using the latest available MGS orbit and Mars rotational model. 
Then, we self-registered the reprocessed MOLA profiles into a self-consistent reference DTM which serves as a static mean surface of Mars. 
The self-registration procedure is needed because even after incorporating the refined auxiliary data in the reprocessing, residual orbit errors, and untreated pointing and timing errors still present.

\subsection{State of the art reprocessing}
\label{sec:mola_rep_subsec}
\quad The MGS trajectory and the Mars IAU2000 rotational model adopted by the MOLA PEDR dataset date back to almost two decades ago, thus we reprocess the entire MOLA dataset using the latest available MGS orbit model \citep{konopliv2006orbit} and an updated Mars rotational model \citep[IAU2015;][]{archinal2018report}. 
The nominal accuracy of this refined MGS orbit model is $\sim$2~m laterally and $\sim$15~cm radially as revealed by an orbital arc overlap analysis \citep{konopliv2006orbit}. 
The right ascension, declination, and prime meridian from the Mars IAU2015 rotational model are known to about 13 milli arcseconds (mas), 8~mas, and 71~mas at the J2000 epoch, respectively \citep{jacobson2018rotation}. 
Approximately, 1~mas corresponds to about 16~mm on the surface of Mars at the equator. 
During the reprocessing, flagged noise returns and shots with missing attitude information are excluded and range corrections due to the detector response and range walk are applied.
The orientation of the spacecraft reference frame with respect to the International Celestial Reference Frame (ICRF) is extracted from "Camera-matrix" Kernels (CKs) in the Spacecraft, Planet, Instrument, Camera-matrix, Events (SPICE) system \citep{ActonSPICE}. 
Besides, timing biases specific to MOLA, including the MOLA internal timing bias of 117.1875~ms and the CK time tag adjustment bias of -1.15~s, were considered \citep{neumann2001cross-over}.
Furthermore, we take into account the pointing aberration induced by special relativity. 
For the nadir-pointing profiles, this factor can lead to lateral shifts of less than 5~m and radial ones of less than 5~cm, whereas these values can be as high as 10~m and 25~m, respectively, for the intermittently acquired profiles with off-nadir pointing angles up to $35\degree$ \citep{xiao2020geolocation}. 
These off-nadir profiles were acquired by commanding MGS to roll off-nadir >10$\degree$ to fill the gaps beyond the orbital limits of $\sim$87$\degree$S/N, and have amounted to more than 100 throughout the mission. 
The self-consistency of the reprocessed MOLA PEDR is slightly better than the original MOLA PEDR dataset, but still much worse than the MOLA PEDR with the global cross-over analysis \citep{neumann2001cross-over}, refer to \ref{sec:cross-over_misfits_reprocessed}.

\begin{figure}[H]
\centering 
\includegraphics[scale=0.45]{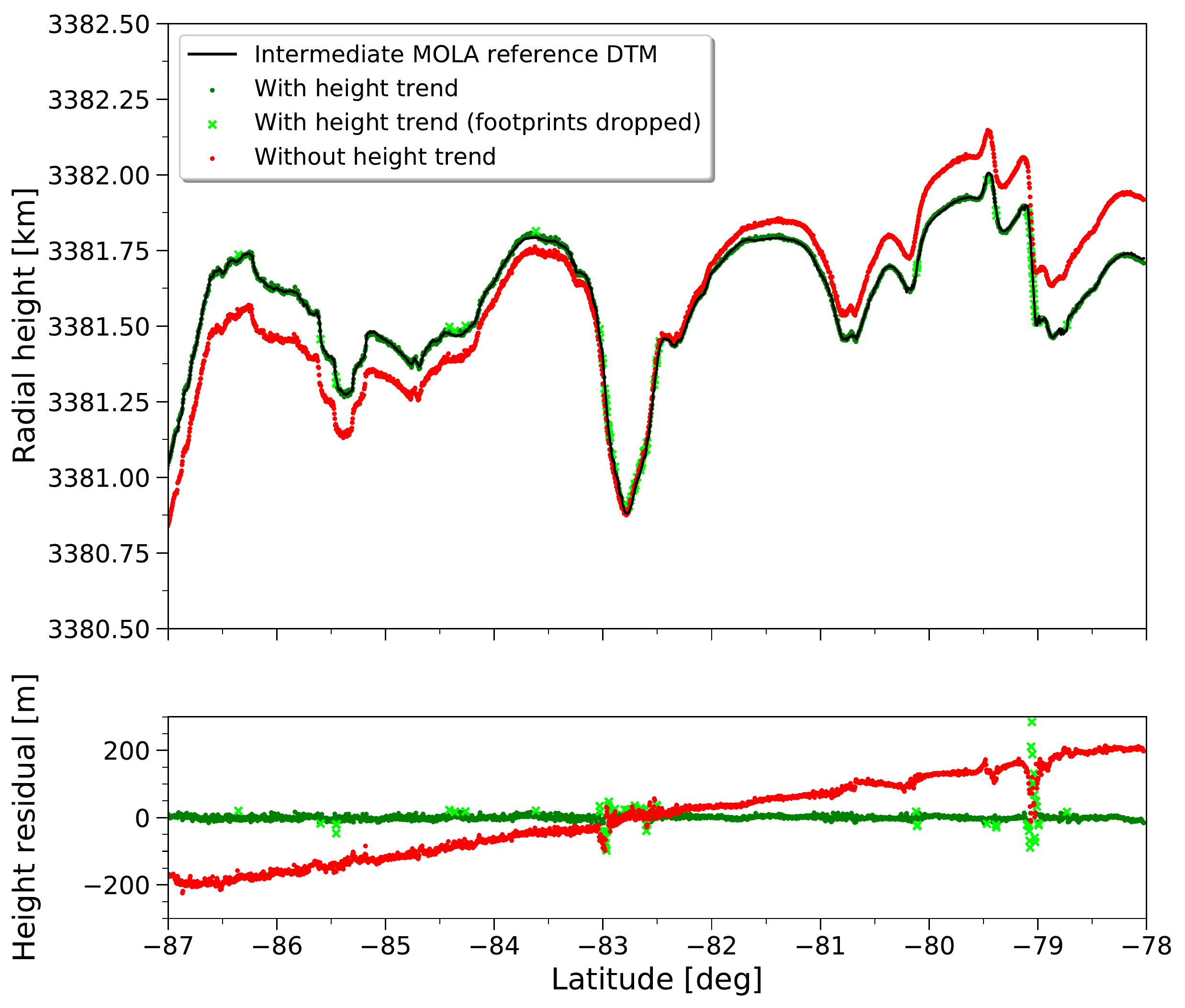}
\caption{Top: Radial heights represented by the intermediate MOLA reference DTM and the footprints along one off-nadir profile (track 14,024) after co-registration. Bottom: height residuals after co-registration. 
Green denotes result with the additional height trend parameter in the co-registration while red is without the additional parameter.}
\label{fig:residual_offnadir}
\end{figure}

\begin{figure}[H]
\centering 
\includegraphics[scale=0.35]{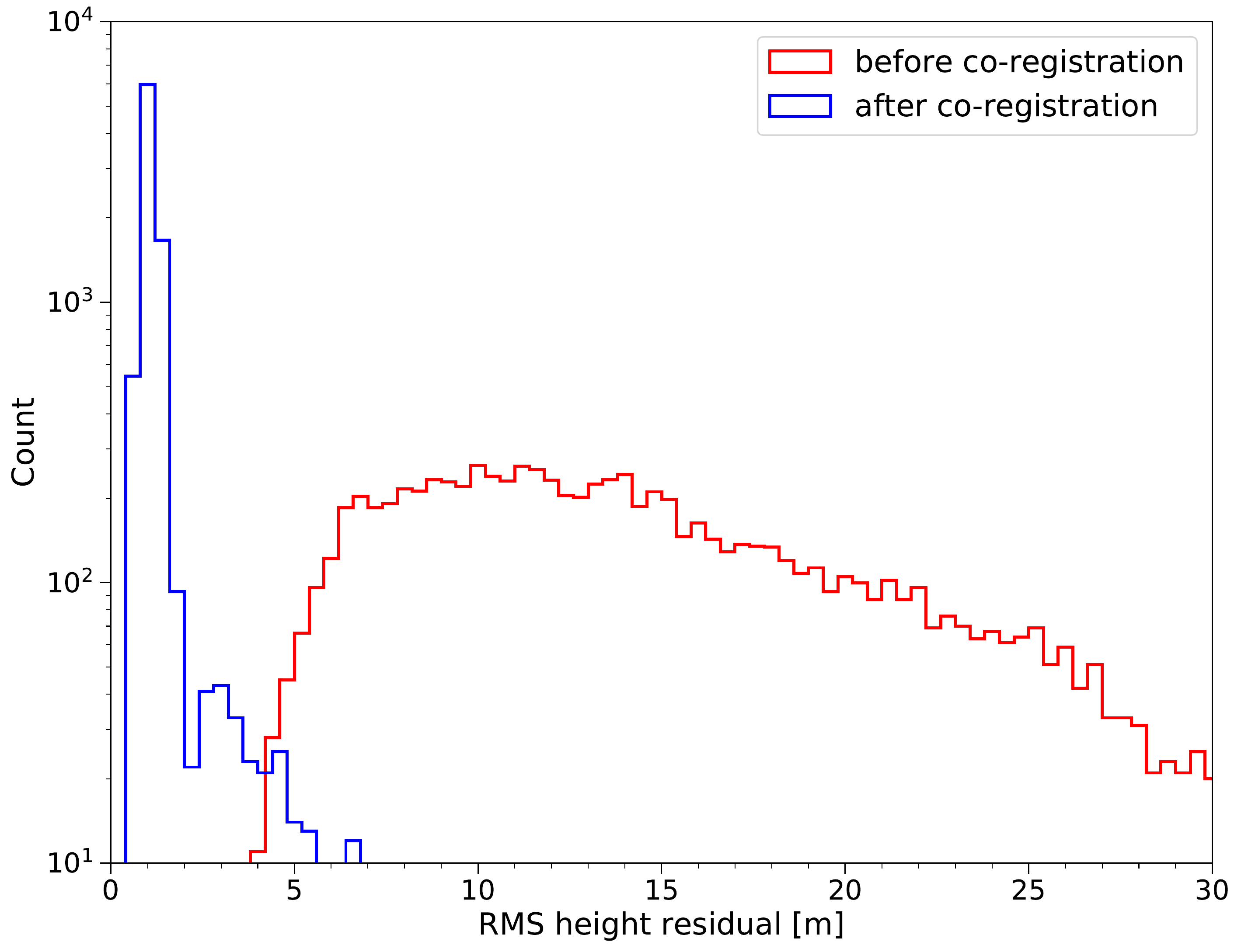}
\caption{Histogram of the RMS height residuals of the MOLA profiles to the intermediate MOLA reference DTM before (red) and after (blue) co-registration.}
\label{fig:MOLA_rms_residuals}
\end{figure}

\begin{figure}[H]
\centering 
\includegraphics[scale=0.9]{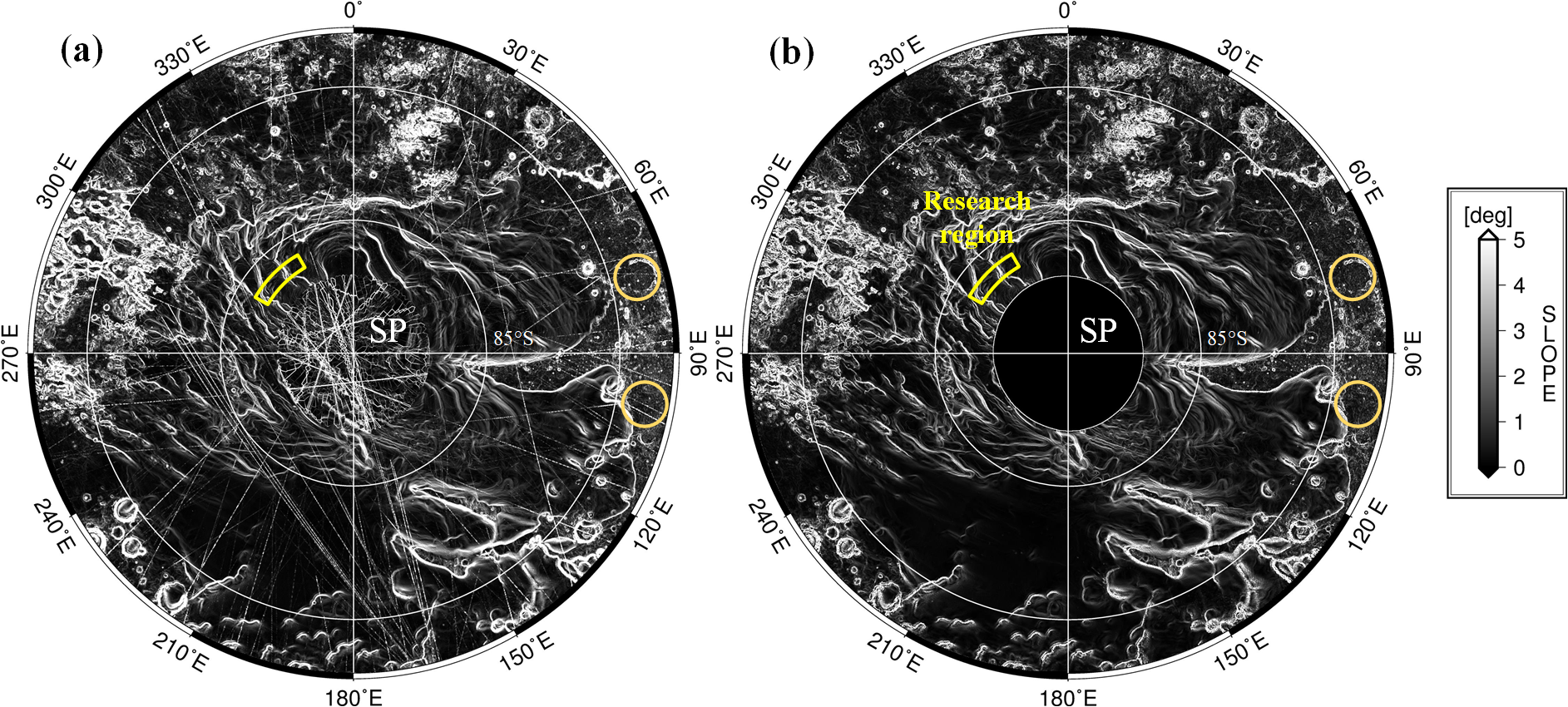}
\caption{(a) Slope map from the reprocessed MOLA profiles before the self-registration (Sec.~\ref{sec:mola_rep_subsec}). 
(b) Slope map from the self-registered MOLA reference DTM. 
The missing coverage poleward of 87$\degree$S is due to off-nadir profiles with RMS height residuals exceeding 2~m being excluded during the self-registration process. 
The fan-shaped region outlined on both plots is the experimental region within annulus 86$\degree$S for the CO$_2$ height variation measurement experiment. 
The adopted projection is polar stereographic centered at the South Pole and the map covers the region from 78$\degree$S to 90$\degree$S. 
The slope was calculated using the $\href{https://gdal.org/programs/gdaldem.html}{gdaldem}$ routine in the Geospatial Data Abstraction Library (GDAL) and the figure is plotted using the Generic Mapping Tool (GMT) software \citep{Wessel2013GMT}.}
\label{fig:MOLA_slope_map}
\end{figure}

\subsection{Self-registration of reprocessed profiles}
\label{sec:self_registration_MOLA}
\quad To remove residual error sources from the MOLA profiles, we follow a two-step strategy in the self-registration process.
First, we select all of the reprocessed nadir-pointing profiles falling within [78$\degree$S,\, 87$\degree$S] and iteratively self-register them to form a consistent intermediate MOLA reference DTM (\ref{sec:self_regist}). 
We set the limit at 78$\degree$S because towards the equator the density of MOLA footprints is gradually decreasing. 
Because of the inclination of the MGS orbit of 87$\degree$S MOLA could not perform accurate measurements in regions close to the pole.
Hence, the footprint density is sparse and their quality is compromised due to off-nadir pointing.
If included, profile segments there can interfere with the self-registration process, thus we set the limit of 87$\degree$S.
In the next step, we co-registered all the reprocessed profiles (including those obtained with off-nadir) falling within [78$\degree$S,\, 90$\degree$S] to the intermediate MOLA reference DTM to form a complete self-registered reference DTM.
In the following we provide more details to the performed steps.

\quad For each iteration in the self-registration in the first step, we select an empirical 25$\%$ of MOLA profiles and then co-register them in map coordinates (\ref{sec:local_co-registration_in_map_coords}) to an intermediate DTM with a spatial resolution of 500~m/pixel constructed from the rest of profiles. 
We iterate this process until no outliers can be discerned by visual inspection after 50 iterations. 
Finally, an intermediate MOLA reference DTM made out of these self-consistent profiles is obtained.

\quad In the second step, we take all individual reprocessed profiles falling within [78$\degree$S,\, 90$\degree$S] and co-register each of them to the intermediate MOLA reference DTM with a parameterization in map coordinates and height trend (\ref{sec:local_co-registration_in_map_coords_and_trend}). 
The additional parameter, height trend, is intended to compensate for a temporal linear tilt observed in a majority of the intermittently-acquired off-nadir profiles. 
Fig.~\ref{fig:residual_offnadir} shows a comparison between parameterizations in map coordinates with and without the height trend for a single off-nadir profile (track 14,024), the extra height trend parameter can absorb much of the observed linear tilt.
This tilt could be induced by attitude errors of the spacecraft. 
Extreme outlier footprints along each of the profiles have been detected by a 3-$\sigma$ filter and excluded during the iterative co-registration process. 
The histogram of the Root Mean Square (RMS) height residuals for all of the profiles before and after the co-registration are shown in Fig.~\ref{fig:MOLA_rms_residuals}. 
The RMS height residuals have generally been reduced from more than 5~m to generally within 2~m after co-registration. 
Due to the bimodal pattern of the histogram after co-registration, we then apply an empirical threshold of 2~m to exclude the long tail that is mainly a result of the off-nadir profiles. In total, 348 out of 8,618 profiles (a ratio of $4\%$) or 1,108,312 out of 36,580,839 footprints (a ratio of $3\%$) have been excluded after this filtering process. 
Unfortunately, no off-nadir profiles covering the area outside the MGS inclination limit (poleward of 87$\degree$S) remain after this filtering. 

\quad Fig.~\ref{fig:MOLA_slope_map} shows slope maps obtained from gridded DTMs from the reprocessed profiles (Sec.~\ref{sec:mola_rep_subsec}) and from the self-registered MOLA reprocessed profiles, erroneous profiles have been removed, individual profiles are tied to each other and speckle noise is dramatically reduced. 
The internal consistency of the reprocessed profiles is much better after this self-registration procedure and even slightly better than the MOLA PEDR with the cross-over corrections \citep{neumann2001cross-over}, refer to \ref{sec:cross-over_misfits_self_registered}. 
The gridded reference DTM (500~m/pixel) from the self-registered MOLA profiles serves as a static reference topography for the dynamic reprocessed MOLA profiles in the local co-registration strategy and post-correction procedure proposed in Sec.~\ref{sec:methods} to obtain the temporal seasonal height variations of the seasonal south polar cap.

\section{Proposed methods for mapping height variations}
\label{sec:methods}

\label{sec:three_methods}
\quad In this section, first, methods for the computation of temporal height differences either at footprints ("F") or cross-overs ("X") will be presented. Each of these two methods is further subdivided into two versions, depending on whether co-registration is applied. For better clarity, those with the co-registration procedure are suffixed with "LC" which stands for "$\textit{local co-registration}$" (see Table \ref{table:methods}). 
The so-called local co-registration means the co-registration is done using "short" intercepted laser profile segments centered either at footprints ("F$\_$LC") or cross-overs ("X$\_$LC"), in comparison to the "long" extending profiles that can stretch thousands of kilometers. 
This local co-registration strategy can better preserve the local information of the seasonal height variations. 
Then, a post-correction procedure to correct for a systematic temporal bias in MOLA heights 
is described. 
Finally, the median-binning of the post-corrected temporal height differences to temporal height variations and the precision quantification are presented. 
The overall flowchart is shown in Fig.~\ref{fig:flowchart}.

\begin{center}
\tablehead{
    \toprule  
    \multirow{2}{*}{Temporal height differences} & \multirow{2}{*}{Count} &
    \multicolumn{2}{c}{Methods} \cr
    \cmidrule(lr){3-4}
    & & \textbf{without} co-registration & \textbf{with} co-registration\cr  
    \midrule} 
    \topcaption[lr]{
    Methods tested and compared in this work. 
    The second column denote the total number of available MOLA footprints or cross-overs in the polar research region, respectively.}
    \begin{supertabular}{p{4.5cm} p{1.8cm} p{3.8cm} p{3.4cm}}
       @ footprints & 194,109 & "\textbf{F}" & "\textbf{F$\_$LC}" \\
       @ cross-overs & 258,069 & "\textbf{X}" & "\textbf{X$\_$LC}" \\
       \bottomrule
    \end{supertabular}
\label{table:methods}
\justifying
\end{center}

\begin{figure}[H]
\centering 
\includegraphics[scale=0.6]{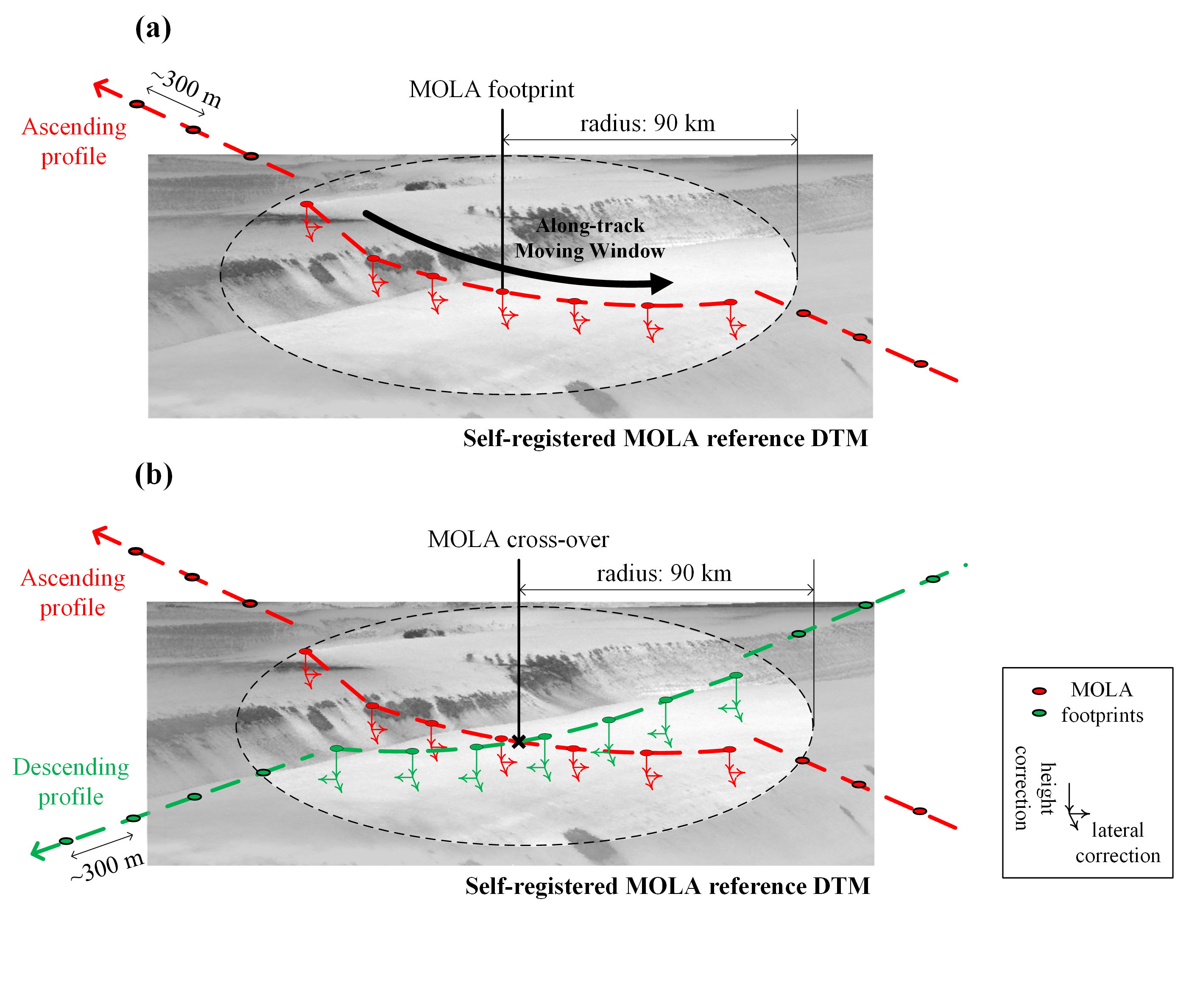}
\caption{Schematic illustration of the proposed methods combined with the local co-registration strategy to detect height change signal of the seasonal south polar cap (example is made on elevated layered terrains). 
(a) "F$\_$LC"-method, the height difference of each MOLA footprint with respect to the self-registered MOLA reference DTM is retrieved from the height correction of the local co-registration process that aligns the local profile segment centered at the footprint to the reference DTM. 
(b) "X$\_$LC"-method, height differences at cross-overs with respect to the self-registered MOLA reference DTM are assigned as the corrections of the local co-registration processes that align local segments of the intersecting profiles centered at the cross-overs to the reference DTM.}
\label{fig:methods}
\end{figure}

\subsection{Obtaining height differences}

\subsubsection{Height differences at footprints}
\quad The "F"-method is straightforward, that is to compute the differences between the heights of the footprints and the corresponding heights of the self-registered MOLA reference DTM at the coincident positions. 
The latter ones are obtained at sub-pixel level through interpolation of neighboring DTM pixels using quadratic splines in both line and sample dimensions. 
We set time stamps of the height differences to be the corresponding acquisition time stamps of the footprints. 
While all of the height information of the footprints in the research area can be utilized, this method is sensitive to the noise inherited in both of the data products, e.g., noisy/unrealistic laser altimeter footprint heights, interpolated pixels of the DTM, and the relative misalignment between them.

\quad This approach can be refined by incorporating the local co-registration strategy, a method we name "F$\_$LC". 
As shown in Fig.~\ref{fig:methods}a, footprints along the profile falling within a circle centered at a specific footprint and with a radius of 90 km are selected to form a local profile segment ($\sim$300 footprints on each side, see also Sec.~\ref{sec:min_points_stability} for the number of footprints needed to stabilize the co-registration). 
Due to the frozen circular orbit of MGS and constant MOLA firing rate (10~Hz), in practice, this is done by including footprints acquired within $\pm$30 seconds of the acquisition time stamp of the footprint in question. 
For nadir-pointing profiles this is equivalent to about 90~km or about 300 footprints on each side. 
Then, this local profile segment is co-registered to the underlying self-registered MOLA reference DTM, parameterized in map coordinates as detailed in \ref{sec:local_co-registration_in_map_coords}. 
Instead of simple height differencing as in the "F"-method, the height correction of the aforementioned local co-registration process is considered as the height difference of this specific footprint. 
We assign the time stamp of this height difference as the reflection time of this specific laser footprint. 
This procedure is repeated for all footprints along each of the profiles that fall within the research region to obtain a series of  height differences for every MOLA footprint.

\subsubsection{Height differences at cross-overs}
\quad The "X"-method is implemented in three steps. 
First, we locate the intersection points, i.e., cross-overs, between each ascending/descending or descending/ascending ground track pair assuming linearity of the local tracks. 
Then, we interpolate the heights at the intersection point from each track and subsequently subtract these from each other. 
Similarly, we interpolate the time stamps at the cross-over for each of the ascending and descending track. 
To treat the data uniformly, each cross-over height difference is counted twice: once with the time stamp of the ascending track, and again, with the time stamp of the descending track with an opposite sign as in \cite{aharonson2004depth}. 
But this simple handling is expected to introduce in more unwanted errors other than that from interpolating and track lateral shifts due to residual orbit, pointing, and timing errors.

\quad By extending the "X"-method by the local co-registration strategy we arrive at the "X$\_$LC"-method (see Fig.~\ref{fig:methods}b). 
Here, we first select local profile segments from intersecting profiles.
The segments are constrained by a circle with a radius of 90~km that is centered at the cross-over point ($\sim$300 footprints on each side).
Then, we co-register the two local profile segments to the underlying self-registered MOLA reference DTM parameterized in map coordinates (\ref{sec:local_co-registration_in_map_coords}). 
At the end, height differences at each of the cross-overs are treated as the two corresponding height corrections out of the co-registration processes, circumventing the interpolation and track lateral shift issues presented in the "X"-method. 
Likewise as in the "X"-method, time tags of the acquired height differences are marked as the corresponding time tags of the cross-overs interpolated from adjacent footprint time stamps.   
This procedure is repeated over all cross-overs within the research region to obtain a series of height differences from cross-overs.

\subsubsection{Implementation of co-registration in map coordinates}
\label{sec:co-reg_MOLA}
\quad The MGS/MOLA orbit, pointing, and timing errors can be amply modeled by harmonic signals that feature both once-per-revolution and twice-per-revolution components \citep{neumann2001cross-over}. 
However, during the short time of a local profile segment (a maximum of 600 footprints, approximately 1 minute in duration), the lateral shifts of these local profile segments can be treated as constants and can then be absorbed by corrections in the line and sample directions during the co-registration process (\ref{sec:local_co-registration_in_map_coords}). 
In addition, we also solve for the height correction which can be interpreted mainly as the seasonal condensation/sublimation signal at the Martian polar regions. 
As such, we do not implement corrections to the two-way range (e.g., uncertainty in range walk correction of approximately 30~cm for unsaturated footprints \citep{abshire2000mars}), since they are highly correlated with the height change signal, especially in MOLA's nadir-pointing configuration.

\quad We use an identity weighting matrix for height differences at footprints' locations in the co-registration.
We exclude outliers that fail the 3-$\sigma$ criterion to eliminate the side impacts of MOLA false and noisy returns, returns from snowflakes, and interpolated DTM pixels. 
The co-registration iteration stops when updates are less than $10^{-3}$ pixel (or 0.5~m) in the line and sample corrections and less than 1~mm in height correction. 
Normally, it takes about 15 iterations to reach these convergence thresholds.
In addition, we also discard measurements from profile segments with RMS height residuals greater than 4~m after co-registration (typical RMS values range from 1.5~m to 3.5~m), or with remaining valid footprints less than 400 (refer to the number of footprints needed to stabilize the co-registration in Sec.~\ref{sec:min_points_stability}).

\begin{landscape}

\begin{figure}[H]
\centering 
\includegraphics[scale=0.52, angle=0]{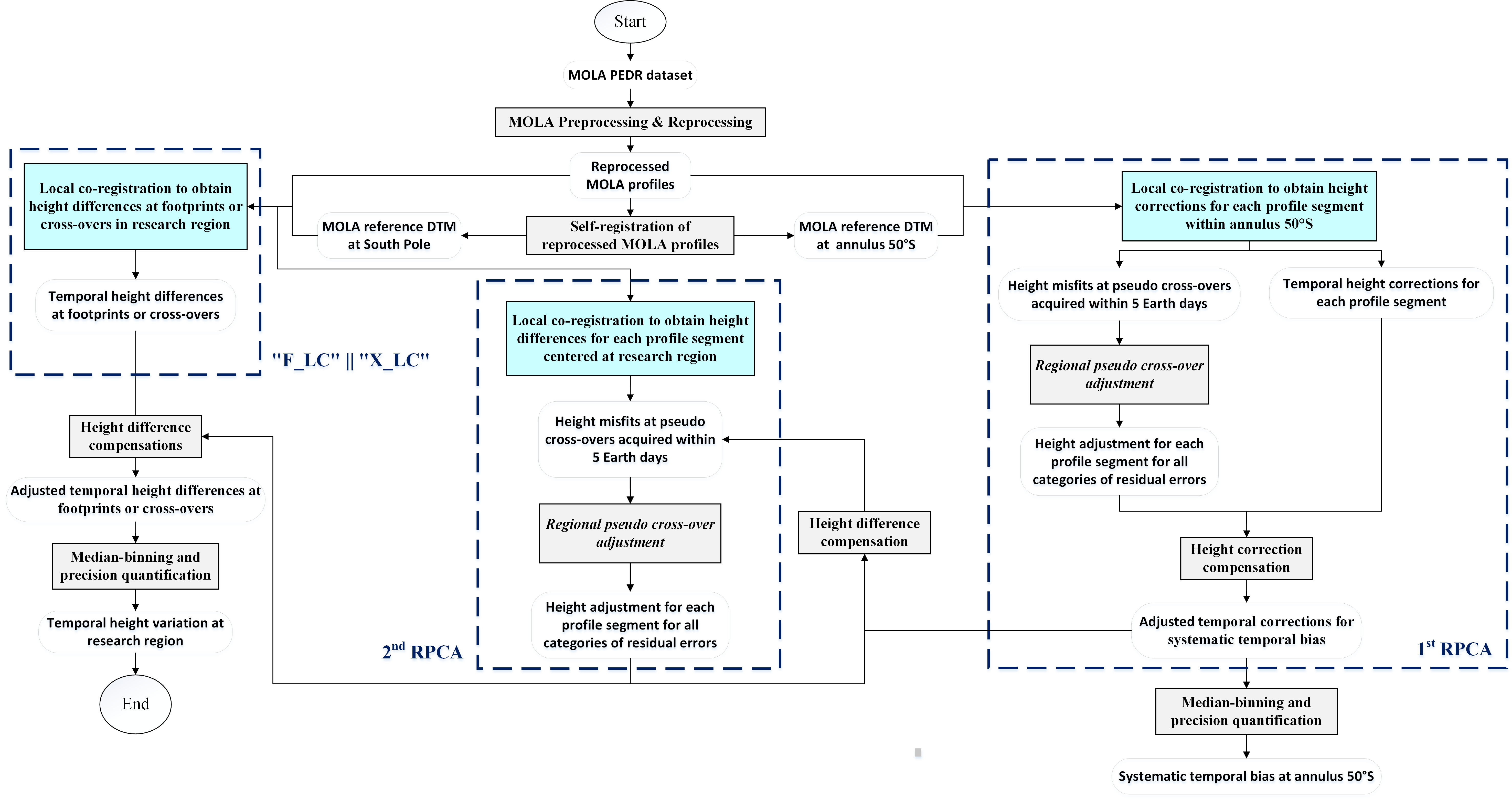}
\caption{Flowchart for measuring the temporal height variations of the Martian seasonal polar caps using "F$\_$LC" or "X$\_$LC"-method post-corrected by bi-RPCA.}
\label{fig:flowchart}
\end{figure}

\end{landscape}

\subsection{Post-correction procedure}
\label{sec:sys_bias_and_bi-RPCA}
\quad We have not observed any temporal trend in the radial differences between the former orbit used in MOLA PEDR and the one by \cite{konopliv2006orbit} used in the reprocessing of MOLA PEDR (see Fig.~\ref{fig:orbit_comparison} in Sec.~\ref{sec:sys_bias_appex}). 
Thus, the global systematic temporal bias in MOLA footprint heights is still inherited in our reprocessed MOLA dataset. 
To correct for this artifact as well as to compensate for residual errors in the reprocessed MOLA profiles and self-registered MOLA reference DTMs as to further improve the precision of the obtained height variations, we propose a novel two-step post-correction procedure as detailed in the following subsections.

\subsubsection{Regional pseudo cross-over adjustment}
\quad The regional cross-over adjustment (RCA) over a relatively small, non-dynamic area has been extensively used to improve the internal consistency of the biased altimetry profiles on Earth \citep{ewert12}. 
Here, we adopt pseudo cross-overs instead of cross-overs in the post-correction process and we perform a regional pseudo cross-over adjustment (RPCA) to improve the internal consistency of the local profile segments. 
Providing the self-registered MOLA reference DTM is available, any particular pair of profile segments forms a so-called DTM-based “pseudo cross-over” \citep{barker2016lola}. 
For example, for the pseudo cross-over formed by the $k^{\rm{th}}$ and $l^{\rm{th}}$ profile segments, the height misfit at the pseudo cross-over ($\Delta h_{\rm{pc}}$) is assigned as
\begin{linenomath*}
\begin{equation}
\Delta h_{\rm{pc}} = \Delta h^{k} - \Delta h^{l}\,,
\label{eq:pseudo_xover_misfit}
\end{equation}
\end{linenomath*}
where $\Delta h^{k}$ and $\Delta h^{l}$ are the height corrections of the $k^{\rm{th}}$ and $l^{\rm{th}}$ profile segments to the reference DTM from the co-registration processes, respectively (Fig.~\ref{fig:pseudo_xover_illustration}). 
Thus, the impacts of the lateral shifts of the profiles and the height interpolation errors in the case of the cross-overs are avoided. 
To minimize the interference of the CO$_2$ height change signal, only pseudo cross-overs acquired within 5 days are used in the analysis. 
Pseudo cross-overs can have great advantages over cross-overs. 
Since the profile segment pairs forming the pseudo cross-overs do not have to actually intersect, the available number of "cross-overs" is  multiplied. 
Furthermore, the profile segment pair forming a pseudo cross-over can be widely distributed across the region, offering "global" constrains on the adjustment. 
In contrast, the profile segment pairs forming cross-overs are spatially confined to be around their intersection points. 
Additionally, the time-consuming process of locating tens of millions of cross-overs can be spared.

\begin{figure}[htb]
\centering
\includegraphics[scale=1.2]{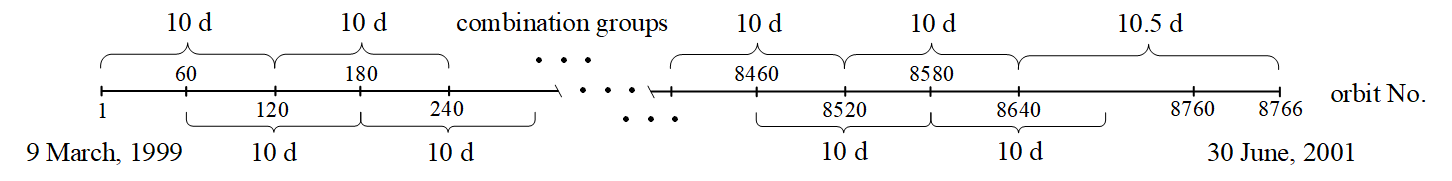}
\caption{Slicing of the MOLA data into combination groups of 10 days to speed up the selection of pseudo cross-overs acquired within 5 days.
}
\label{fig:pseudo_xover_forming}
\end{figure}
\quad For the identification of pseudo cross-overs acquired within 5 days, pure combinations of all the available profile segments can lead to tremendous number of candidate pseudo cross-overs, especially at entire annuli are considered. 
This can dramatically slow down the subsequent filtering process based on time interval criterion of 5 days. 
To overcome this, we propose a divide-and-conquer algorithm to speed up the process (Fig.~\ref{fig:pseudo_xover_forming}). 
The proposed strategy is to slice the time-sorted profiles into each of the 72$\times$2 consecutive combination groups that span 10 days each (plus one that span 10.5 days at the end), and retrieve all the possible combinations of profile segment pairs within each group. 
Then, we exclude combinations with orbit number difference greater than 60 (equals to 5 days) and remove the duplicates.

\begin{figure}[htb]
\centering
\includegraphics[scale=1]{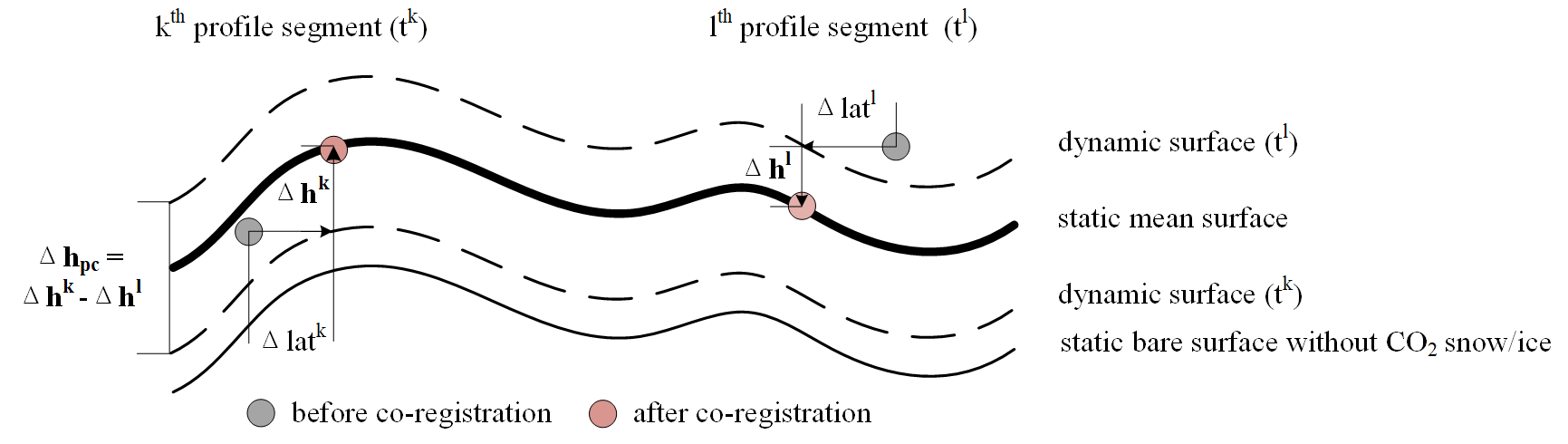}
\caption{Illustration of the pseudo cross-over formation. 
The static mean surface is provided by the self-registered MOLA reference DTM. 
The surface height misfit at the pseudo cross-over $\Delta h_{\rm{pc}}$, when time interval between $t^k$ and $t^l$ is less than 5 days, is obtained from the difference in the height corrections in the co-registration processes of the two profile segments to the static mean surface. 
$\Delta \rm{lat}^{k}$ and $\Delta \rm{lat}^{l}$ are the lateral corrections of the profile segments in the co-registration processes.}
\label{fig:pseudo_xover_illustration}
\end{figure}

\quad We assume a constant height correction for each of the profile segment and the inversion is solved by a linear least-squares adjustment. 
Ridge regression is introduced to stabilize the ill-posed inverse problem due to correlations between the model parameters, which are constant height corrections for each of the profile segments. 
The cost function to be minimized is
\begin{linenomath*}
\begin{equation}
\arg \min_{\hat{\mathbf{x}}_\alpha} \, (\mathbf{Ax}-\mathbf{b})^{\rm{T}}(\mathbf{Ax}-\mathbf{b})+\alpha \mathbf{x}^{\rm{T}}\mathbf{x} \,,
\label{eq:cost_cundtion}
\end{equation}
\end{linenomath*}
where $\mathbf{x}$ denotes vector of the model parameters. 
$\mathbf{A}$ is the design matrix which consists of elements of 0, 1 and -1, and $\mathbf{b}$ is the vector of the observables, i.e. the height misfits at the pseudo cross-overs acquired within 5 days as formulated in Eq.~(\ref{eq:pseudo_xover_misfit}). 
The second term in Eq.~(\ref{eq:cost_cundtion}) is the introduced regularization and $\alpha$ is the regularization strength. 
The vector of $\hat{\mathbf{x}}_\alpha$ denotes the regularized parameter solution \citep{xu92}
\begin{linenomath*}
\begin{equation}
\hat{\mathbf{x}}_\alpha=\left(\mathbf{A}^{\rm{T}} \mathbf{A}+\alpha \mathbf{I}\right)^{-1} \mathbf{A}^{\rm{T}}\mathbf{b} \,,
\label{eq:reg_solution}
\end{equation}
\end{linenomath*}
where $\mathbf{N}=\mathbf{A}^{\rm{T}}\mathbf{A}$ is the diagonally banded normal equation matrix and $\mathbf{A}^{\rm{T}} \mathbf{A}+\alpha \mathbf{I}$ the normal equation matrix with regularization. 
Ridge regression is a biased estimation and the induced biases on the inverted height corrections are
\begin{linenomath*}
\begin{equation}
\rm{bias}(\hat{\mathbf{x}}_\alpha)\approx-\alpha \left(\mathbf{A}^{\rm{T}} \mathbf{A}+\alpha \mathbf{I}\right)^{-1} \hat{\mathbf{x}}_\alpha \,.
\label{eq:reg_bias}
\end{equation}
\end{linenomath*}
We partially correct for the biases in Eq.~(\ref{eq:reg_bias}) to the regularized parameter solution to guarantee a further reduction in terms of the Mean Squared Error (MSE) criterion as in \cite{shen12}
\begin{linenomath*}
\begin{equation}
\hat{\mathbf{x}}_c=\left(\mathbf{I}+\alpha \sum_{i} \frac{\mathbf{u}_{i} \mathbf{u}_{i}^{\mathrm{T}}}{\lambda_{i}+\alpha}\right) \hat{\mathbf{x}}_\alpha
\,\,\,\,\, \mathrm{s.t.} \,\,\,
\left\{\begin{array}{l}
\lambda_i < \alpha \\
\left(2 \lambda_{i}+3 \alpha\right) \operatorname{tr}\left(\mathbf{D}_{\alpha} \mathbf{u}_{i} \mathbf{u}_{i}^{\mathrm{T}}\right) <\alpha \left(\mathbf{u}_{i}\hat{\mathbf{x}}_\alpha\right)^{2}
\end{array}\right. \,,
\label{eq:partially_compensate_reg_bias}
\end{equation}
\end{linenomath*}
where $\hat{\mathbf{x}}_c$ is the vector of the applied partial corrections, $\mathbf{u}_{i}$ is the $i^{\mathrm{th}}$ column vector corresponding to the $i^{\mathrm{th}}$ eigenvalue in the spectral decomposition of $\mathbf{N}$, and $\mathbf{D}_{\alpha}$ is the variance–covariance matrix of the estimated parameters $\hat{\mathbf{x}}_\alpha$ \citep{shen12}.

\quad We adopt the L-curve graphical tool to pin down an initial guess of the optimal regularization strength \citep{hansen98}. 
The relations of the regularization-induced biases, scaled median absolute deviation ($\scli{\mathrm{\small MAD}}{s}$) of the derived temporal variation and the the variance inflation factor (VIF) with respect to the regularization strength serve as additional considerations. 
The VIF measures the multicolinearity and possible aliasing among model parameters and should be less than 5 for sufficient orthogonality \citep{rogerson19}. 
In addition, the evolution of the peak-to-peak magnitude of the temporal height variation with respect to the regularization strength is inspected. 
The ultimate optimal regularization strength is decided as the critical point when the aforementioned peak-to-peak variation magnitude begins to stabilize at its maximum.

\quad To take advantage of the high sparsity as well as the symmetric positive-definite nature of the normal equation matrix with regularization, the inversion is carried out using the conjugate gradient method which is an iterative Krylov subspace method intended specifically for large symmetric positive-definite matrix \citep{shewchuk1994CG}. 
Since regularization is applied in the inversion, no preconditioner is introduced in the conjugate gradient method to accelerate its convergence. 
To ensure the correctness of this iterative method, we also carry out the inversion using the Unsymmetric MultiFrontal PACKage (UMFPACK) direct sparse solver \citep{davis2004sparse} for cross-validation purpose. 
Here, the Cholesky factorization of the normal equation matrix with regularization is used in order to speed up the computation and to stabilize the inversion, which is more time-efficient with highly sparse matrix than QR factorization.

\subsubsection{Post-correction procedure implementation}
\label{sec:bi-RPCA}
\begin{figure}[H]
\centering
\includegraphics[scale=1.2]{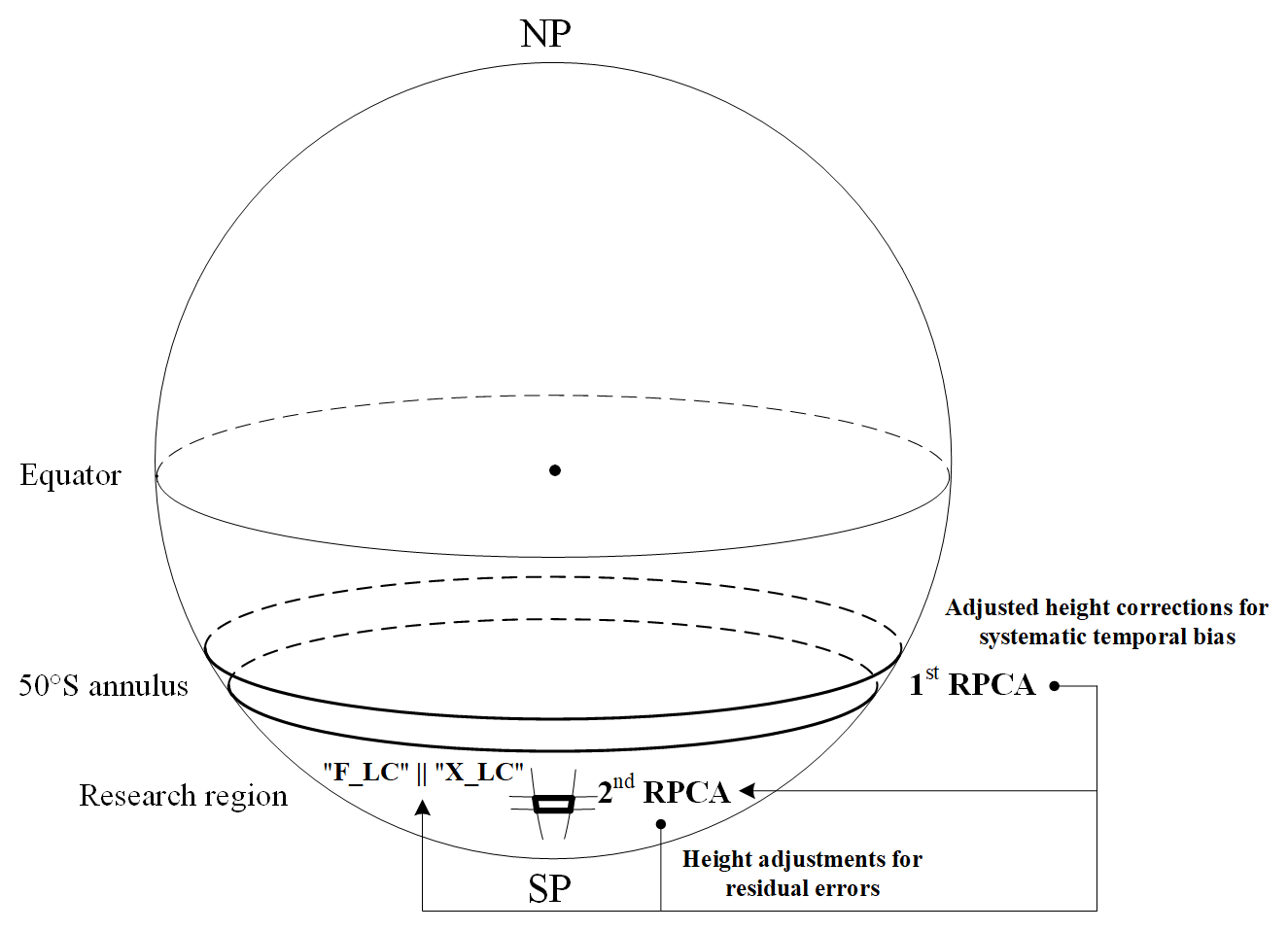}
\caption{Illustration of the bi-RPCA post-correction procedure. 
The 1$^{\rm{st}}$ RPCA is performed at annulus 50$\degree$S to obtain the adjusted height corrections for the systematic temporal bias which are then used to correct for profile segments at the polar regions. 
The 2$^{\rm{nd}}$ RPCA is carried out in the research region within a polar annulus to improve the precision of the obtained temporal height variation using "F$\_$LC" or "X$\_$LC"-method.}
\label{fig:bi-RPCA}
\end{figure}

\quad To correct for the global systematic temporal artifact and residual errors in the reprocessed MOLA profiles and self-registered MOLA reference DTMs, we propose a procedure based on two consecutive RPCAs as "bi-RPCA" that links the polar regions with seasonal height variation and annulus 50$\degree$S of [44$\degree$S,\, 56$\degree$S] without height variation \citep{aharonson2004depth, piqueux15}. 
The schematic plot of bi-RPCA is shown in Fig.~\ref{fig:bi-RPCA} and the main steps are as follows: 

\quad \textbf{Step 1}: Self-registration of MOLA profiles parameterized in map coordinates in annulus 50$\degree$S to set up a reference DTM. 
This reference DTM can serve the purpose to correct for the global systematic temporal bias; 

\quad \textbf{Step 2}: We first obtain height corrections $\Delta h_{\rm{bias}}^{k}$ and $\Delta h_{\rm{bias}}^{l}$ for the $k^{\rm{th}}$ and $l^{\rm{th}}$ profile segments for the systematic temporal bias by co-registering these two profile segments to the reference DTM parameterized in map coordinates. 
Then, the height misfit of the $i^{\rm{th}}$ pseudo cross-over that is formed by these two profile segments is assigned as
\begin{linenomath*}
\begin{equation}
\Delta h_{\rm{pc}}^{i} = \Delta h_{\rm{bias}}^{k} - \Delta h_{\rm{bias}}^{l} \,.
\end{equation}
\end{linenomath*}
Subsequently, using these misfits as the observables, we apply the 1$^{\rm{st}}$ RPCA to acquire the height adjustments for each of these height corrections to compensate for residual errors in the reprocessed MOLA profiles and the MOLA reference DTM. 
Assuming the height adjustments for $k^{\rm{th}}$ and $l^{\rm{th}}$ profile segments are $\Delta h_{\rm{bias\_adj}}^{k}$ and $\Delta h_{\rm{bias\_adj}}^{l}$, then the 1$^{\rm{st}}$ RPCA adjusted height corrections for the systematic temporal bias as to these two profile segments are given by

\begin{linenomath*}
\begin{equation}
\begin{array}{l}
\Delta h_{\rm{bias\_adjed}}^{k} = \Delta h_{\rm{bias}}^{k} - \Delta h_{\rm{bias\_adj}}^{k} \\
\Delta h_{\rm{bias\_adjed}}^{l} = \Delta h_{\rm{bias}}^{l} - \Delta h_{\rm{bias\_adj}}^{l}
\end{array} 
\,;
\label{eq:post_correcred_heigh_corrections_S50}
\end{equation}
\end{linenomath*}

\quad \textbf{Step 3}: Performing the 2$^{\rm{nd}}$ RPCA at the research region using pseudo cross-overs acquired within 5 days. 
Considering the most rapid height growth rate of $\sim$2.5~m in 60 days during the accumulation process in the southern winter \citep{aharonson2004depth}, the maximum real height variation within 5 days is just $\sim$0.2~m. 
For the $j^{\rm{th}}$ pseudo cross-over formed by the $k^{\rm{th}}$ and $l^{\rm{th}}$ profile segments, we first obtain height corrections $\Delta h_{\rm{hc}}^{k}$ and $\Delta h_{\rm{hc}}^{l}$ by co-registering these two profile segments to the self-registered MOLA reference DTM in the South Pole (Sec.~\ref{sec:mola_rep}) parameterized in map coordinates. 
If the length of the profile segments within the research region is short, we extend the profile segments on both sides to be 600 footprints in total to stabilize the co-registration. 
Assuming that the systematic temporal bias at the south polar regions equals that obtained at annulus 50$\degree$S, the height misfit of the $j^{\rm{th}}$ pseudo cross-over after applying the height corrections for the systematic temporal bias from \textbf{Step 2} is
\begin{linenomath*}
\begin{equation}
\Delta h_{\rm{pc}}^{j} = \Delta h_{\rm{hc}}^{k} - \Delta h_{\rm{bias}}^{k} + \Delta h_{\rm{bias\_adj}}^{k} - (\Delta h_{\rm{hc}}^{l} - \Delta h_{\rm{bias}}^{l} + \Delta h_{\rm{bias\_adj}}^{l}) \,.
\end{equation}
\end{linenomath*}
The obtained height adjustments from the 2$^{\rm{nd}}$ RPCA process for residual errors of the $k^{\rm{th}}$ and $l^{\rm{th}}$ profiles are denoted as $\Delta h_{\rm{hc\_adj}}^{k}$ and $\Delta h_{\rm{hc\_adj}}^{l}$, respectively. 
These height adjustments are due to residual errors in the reprocessed MOLA profiles, the self-registered MOLA reference DTM as well as the height corrections for the systematic temporal bias from \textbf{Step 2}. 
In particular, the self-registration to make the reference DTM at the Martian South Pole can be problematic. 
The assumption that the offsets for each of the profiles are 3D constants works well when the shape of the topography stays unchanged, but not any more when spatially and temporally inhomogeneous height variations are present as is the case at the Martian poles; 

\quad \textbf{Step 4}: We denote the height differences from "F$\_$LC"-method for the $m^{\rm{th}}$ and $n^{\rm{th}}$ footprints as $\Delta h_{\rm{LC}}^{k, m}$ and those from "X$\_$LC" for cross-overs along the $k^{\rm{th}}$ and $l^{\rm{th}}$ profile segments as $\Delta h_{\rm{LC}}^{l, n}$.
The corresponding adjusted "F$\_$LC" and "X$\_$LC" height differences after the bi-RPCA post-correction for systematic temporal bias in \textbf{Step 2} and residual errors in \textbf{Step 3} are  
\begin{linenomath*}
\begin{equation}
\begin{array}{l}
\Delta h_{\rm{bi-RPCA}}^{k, m} = \Delta h_{\rm{LC}}^{k, m} - \Delta h_{\rm{bias}}^{k} + \Delta h_{\rm{bias\_adj}}^{k} - \Delta h_{\rm{hc\_adj}}^{k} \\
\Delta h_{\rm{bi-RPCA}}^{l, n} = \Delta h_{\rm{LC}}^{l, n} - \Delta h_{\rm{bias}}^{l} + \Delta h_{\rm{bias\_adj}}^{l} - \Delta h_{\rm{hc\_adj}}^{l}
\end{array}
\,.
\label{eq:post_correcred_heigh_corrections}
\end{equation}
\end{linenomath*}

\subsection{Median-binning and precision quantification}
\label{sec:median-binning}
\quad  After obtaining the bi-RPCA post-corrected height differences on either footprints or cross-overs from Eq.~(\ref{eq:post_correcred_heigh_corrections}), these observables will then be median-binned against their corresponding time stamps to form a height variation time series. 
First, we evenly split the whole observation period from March 1999 to June 2001 into 120 temporal bins of $\sim$5 days in duration each, and then sort out the obtained height differences into their respective temporal bins. 
An analysis of the height differences within each temporal bin shows that the distribution is not Gaussian, thus the median-binning approach that computes the median of all observables in each bin is adopted. 
Beforehand, we apply an iterative 2.5-$\sigma$ filter to remove outliers within each temporal bin in which approximately $5\%$ on average of the observables can be excluded. 
The precision of the computed median within each temporal bin is represented by a scaled median absolute deviation (MAD) \citep{leys2013mad}. 
The MAD is defined as the median of the absolute deviations from the samples' median:
\begin{linenomath*}
\begin{equation}
   \mathrm{\small MAD} =      \operatorname{Md}\left[|\Delta\mathbf{h}-\operatorname{Md}\left[\Delta\mathbf{h}\right]|\right]\,,
\end{equation}
\end{linenomath*}
where $\Delta\mathbf{h}$ denotes the sample vector which contains all the bi-RPCA post-corrected height differences within each temporal bin and $\operatorname{Md}\left[\,\right]$ is the median operator. 
In order to use the MAD as a consistent estimator similar to the standard deviation of a Gaussian distribution, a scaled MAD is adopted instead
\begin{linenomath*}
\begin{equation}
   \scli{\mathrm{\small MAD}}{s} = k\mathrm{\small MAD}\,,
\end{equation}
\end{linenomath*}
where k = 1.4826 is the scale factor \citep{rousseeuw1993median}.

\section{Results}
\label{sec:res}

\quad We focus our analysis on the research region of [85.75$\degree$S, 86.25$\degree$S, 300$\degree$E, 330$\degree$E] within annulus 86$\degree$S near the MGS inclination limit ($\sim$ 87$\degree$S) where the density of footprints and the number of cross-overs maximizes (see Fig.~\ref{fig:MOLA_slope_map} for location). 
This region is also chosen due to its location right on the residual polar cap of the South Pole where maximum seasonal height variation due to the CO$_2$ condensation/sublimation can be expected \citep{aharonson2004depth}.

\subsection{Systematic temporal bias obtained at annulus 50$\degree$S}
\label{sec:RPCA_50S}

\begin{figure}[H]
\centering
\includegraphics[scale=0.35]{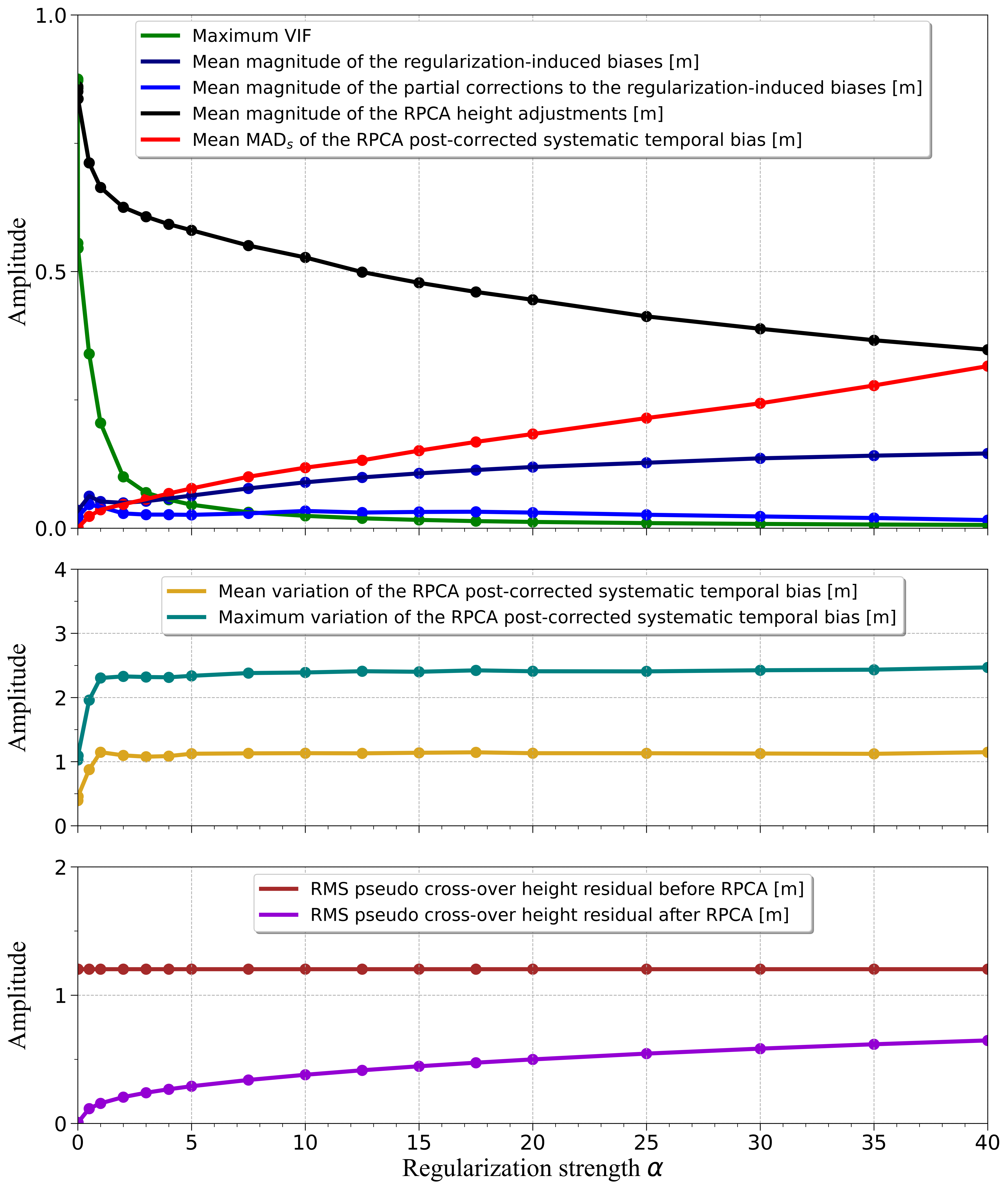}
\caption{Ridge trace for the 1$^{\rm{st}}$ RPCA regularized inversion at annulus 50$\degree$S.}
\label{fig:ridge_trace_50S}
\end{figure}

\quad For the 1$^{\rm{st}}$ RPCA of the bi-RPCA procedure performed at annulus 50$\degree$S, 8,332 out of 8,766 MOLA profile segments are available after the local co-registration.
Thus, 434 profile segments were excluded because they have not reached the threshold for the RMS of height residuals or the minimum number of remaining footprints (see section~\ref{sec:co-reg_MOLA}).
The pseudo cross-overs acquired within 5 days amount to 456,757 and are used in the least squares inversion. 
Sparsity of the normal equation matrix with regularization is 99.1$\%$. The largest and smallest singular values of the super-large sparse design matrix are estimated using the implicitly restarted Arnoldi iteration method \citep{lehoucq1996deflation}. 
The approximated condition number, i.e the ratio of the largest and smallest singular value, is 111, indicating the mild ill-posedness of the inversion model. 

\quad The L-curve indicates an initial guess regularization strength of $\alpha=37.5$.
Then, we take advantage of the ridge trace analysis to pin down the optimal regularization strength (Fig.~\ref{fig:ridge_trace_50S}). 
The maximum VIF of the model parameters become well under 1 when the regularization strength is greater than 0.5. 
Regularization-induced biases of the model parameters generally increase with the regularization, but stay within 0.1~m when the regularization strength is less than 5. 
The variation of the derived systematic temporal bias starts to stabilize at a maximum of $\sim$2.4~m and a mean of 1.1~m when regularization strength approaches 1. 
Thus, the optimal regularization strength of 1 is introduced to stabilize the inversion in the 1$^{\rm{st}}$ RPCA. In this case, a majority of the regularization-induced parameter biases as expressed by Eq.~(\ref{eq:reg_bias}) have been compensated using Eq.~(\ref{eq:partially_compensate_reg_bias}). 

\begin{figure}[H]
\centering
\includegraphics[scale=1]{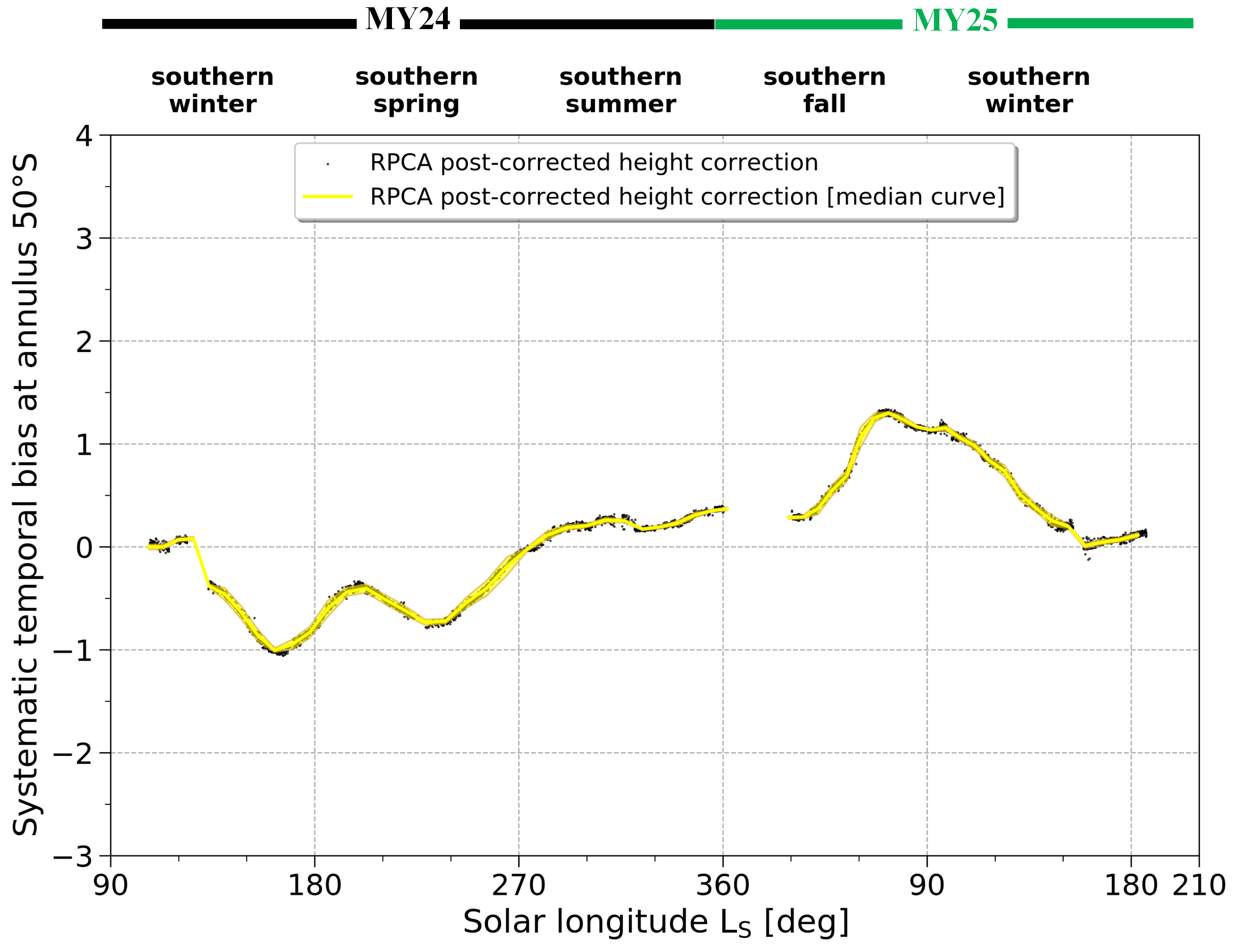}
\caption{Systematic temporal bias derived from the 1$^{\rm{st}}$ RPCA at annulus 50$\degree$S. 
Each black dot in the background represents a RPCA post-corrected height correction from the local co-registration of a specific profile segment within annulus 50$\degree$S to the reference DTM.}
\label{fig:sys_bias_RPCA}
\end{figure}

\quad After the 1$^{\rm{st}}$ RPCA, the RMS of the pseudo cross-over height residuals has decreased from 1.2~m to 0.16~m. 
The mean height correction adjustments for each of the profile segments is 0.66~m. 
The obtained systematic temporal bias by median-binning the RPCA post-corrected height corrections is shown in Fig.~\ref{fig:sys_bias_RPCA}. 
The mean precision as measured by $\scli{\mathrm{\small MAD}}{s}$ is just 3.6~cm, a significantly better value compared to $\sim$50~cm which we obtain by directly median-binning the temporal height corrections from the global cross-over adjustment by \cite{neumann2001cross-over}.
The bias at annulus 50$\degree$S features a minimum of $\sim$-1~m over late southern winter and a maximum of $\sim$1.4~m over late southern fall. 
For cross-validation purpose, we performed the 1$^{\rm{st}}$ RPCA inversion using the UMFPACK direct solver and obtained exactly the same result.

\subsection{Height change time series at research region}

\begin{figure}[H]
\centering
\includegraphics[scale=0.35]{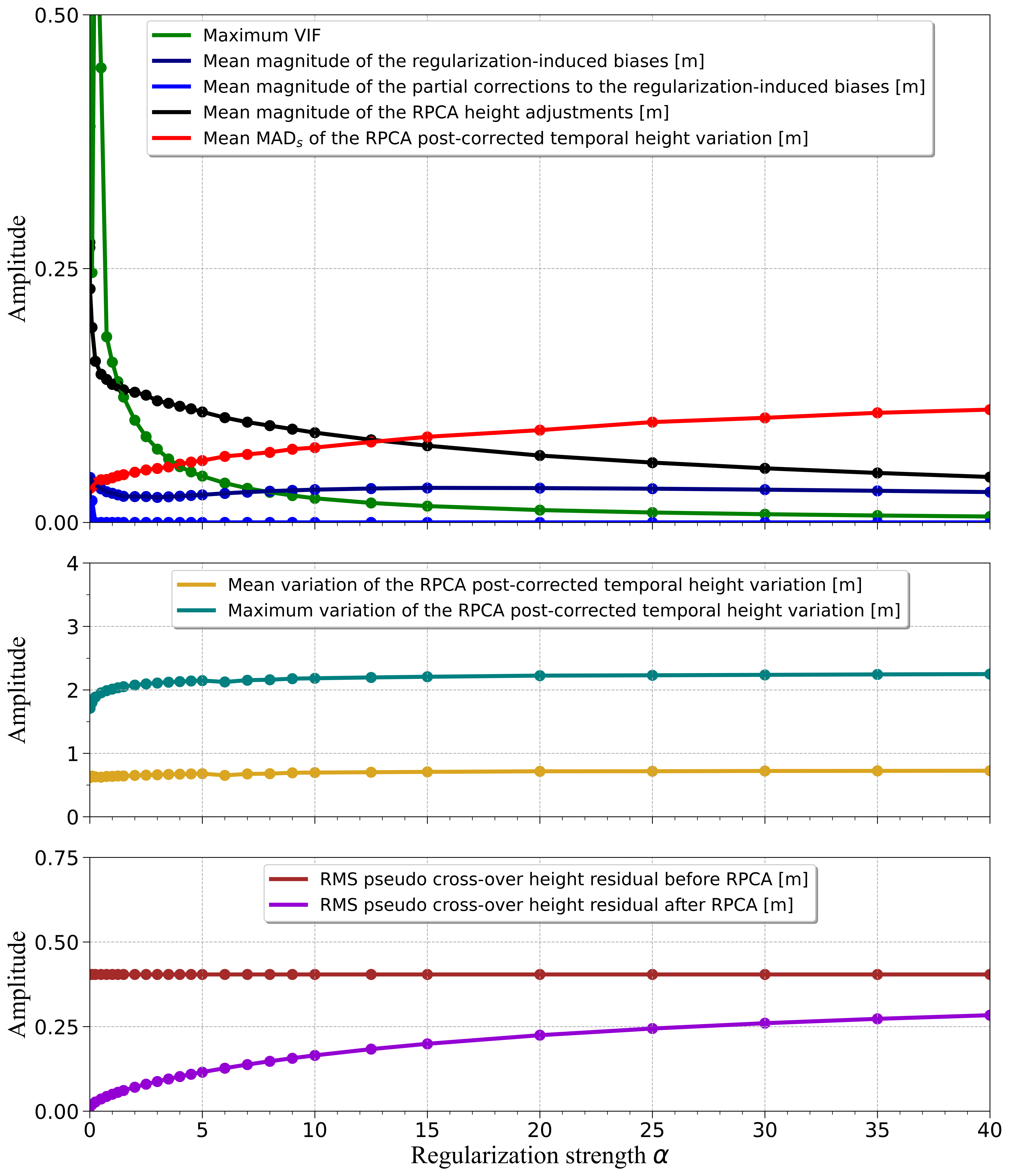}
\caption{Ridge trace for the 2$^{\rm{nd}}$ RPCA regularized inversion at the research region.}
\label{fig:ridge_trace_bin_region}
\end{figure}

\quad Subsequently, the 2$^{\rm{nd}}$ RPCA of the bi-RPCA procedure is then performed at the research region. A total of 1740 profiles are initially available, 1,692 remain after the local co-registration processes, resulting in 18,699 pseudo cross-overs acquired within 5 days. 
The condition number of the design matrix is computed to be as large as 2.4$\times$10$^{17}$, indicating a serious ill-posedness of the inversion model. 
Sparsity of the normal equation matrix with regularization in this case is 98.8$\%$. 

\quad As for the regularization strength, the L-curve suggests an initial guess of $\alpha=16.1$.
Then, we proceed with the ridge trace to determine the optimal regularization strength (Fig.~\ref{fig:ridge_trace_bin_region}). 
The maximum VIF of the model parameters decreases to less than 0.25 once the regularization strength is greater than 1. 
Mean regularization-induced bias of the model parameters stays within 4~cm for the whole range of regularization strengths investigated. 
The maximum variation of the derived seasonal height variation stabilizes at $\sim$2.08~m when regularization strength exceeds 2. 
Thus, the optimal regularization strength of 2 is adopted in the inversion. 
Unfortunately, no components of the regularization-induced parameter biases as expressed by Eq.~(\ref{eq:reg_bias}) have been compensated using Eq.~(\ref{eq:partially_compensate_reg_bias}). However, since these parameters biases are quite small, their impacts on the obtained seasonal height variation measurements are limited. 
The RMS pseudo cross-over height residual has experienced a sixfold decrease from 40~cm to 7~cm once the RPCA is completed. 
The mean inverted height difference adjustments for each of the profile segments stands at 13~cm. The 2$^{\rm{nd}}$ RPCA inversion using the UMFPACK direct solver is also implemented for cross-validation and has led to identical result.

\begin{figure}[H]
\centering
\includegraphics[scale=0.3]{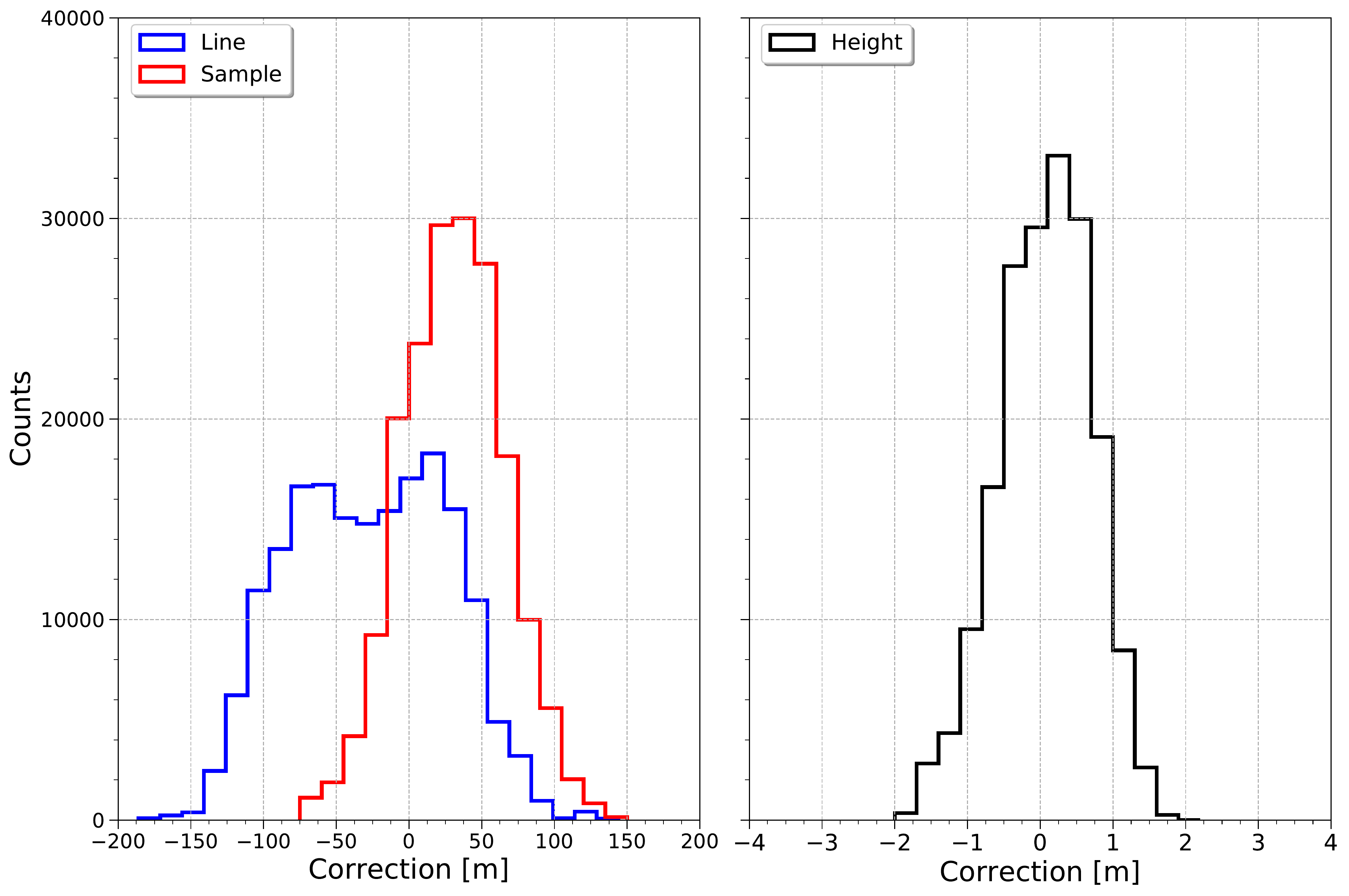}
\caption{Histograms of the line, sample, and height corrections from the local co-registration processes of "F$\_$LC"-method.}
\label{fig:line_sample_height_correction}
\end{figure}

\quad A total of 194,109 footprints are initially available from 1,740 profile segments passing through the research region, 180,254 and 182,514 height differences are obtained after the filtering with the "F" and "F$\_$LC" methods, respectively (refer to Secs.~\ref{sec:co-reg_MOLA} and \ref{sec:median-binning} for the filtering processes). 
For the "F$\_$LC"-method, the histograms of the line, sample, and height corrections from the local co-registration processes are shown in Fig.~\ref{fig:line_sample_height_correction}. 
The line corrections feature two peaks with a median of -26.9~m and a $\scli{\mathrm{\small MAD}}{s}$ of 64.5~m while that of the sample corrections follow a Gaussian-like distribution with a median of 31.1~m and a $\scli{\mathrm{\small MAD}}{s}$ of 36.0~m. 
These figures represent the relative lateral shifts of the reprocessed MOLA profiles. 
The accuracy of the line and sample corrections is better than 5~m.
As to the height corrections, they feature a median of 0.11~m with a $\scli{\mathrm{\small MAD}}{s}$ of 0.66~m, with a formal accuracy of $\sim$10~cm.

\begin{figure}[H]
\centering
\includegraphics[scale=0.6]{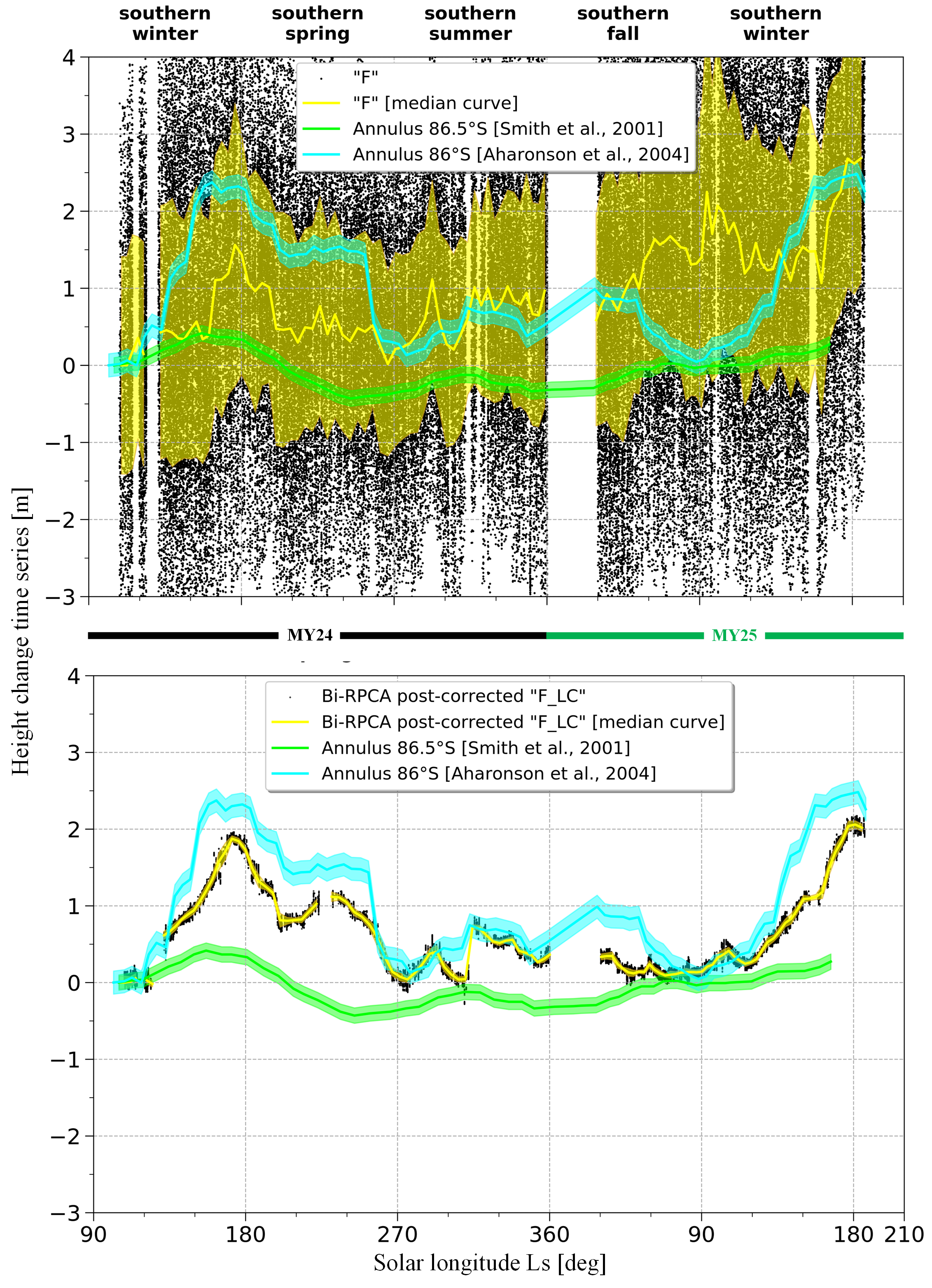}
\caption{Temporal height variation at the research region derived with the "F"-method (top) and the "F$\_$LC"-method post-corrected by the bi-RPCA procedure (bottom).
Yellow shades denote the precision measured by $\scli{\mathrm{\small MAD}}{s}$. 
The previous results obtained with MOLA by \cite{smith2001co2} at annulus 86$\degree$S (green line) and \cite{aharonson2004depth} at annulus 86.5$\degree$S (blue line) are also plotted for comparison.}
\label{fig:height_change_bin_region_F}
\end{figure}

\quad The height variations with respect to solar longitude, $\sclii{L}{s}$, from "F" and  "F$\_$LC" methods, i.e. temporal height differences directly at footprints, are shown in Fig.~\ref{fig:height_change_bin_region_F} (top panel). 
It should be noted that the result form the "F$\_$LC"-method has also been post-corrected by the bi-RPCA procedure. 
The measurements performed with the "F"-method scatter significantly with a $\scli{\mathrm{\small MAD}}{s}$ of $\sim$1.5~m. 
This low precision is a combined effect of errors inherited both in the reprocessed MOLA footprints and the self-registered MOLA reference DTM. 
In addition, the temporal median curve is unbalanced and there is no clear repetition of the curve in the southern winters of MY24 and MY25. 
Significant improvements can be seen after incorporating the local co-registration strategy using the "F$\_$LC"-method and the bi-RPCA post-correction procedure, see Fig.~\ref{fig:height_change_bin_region_F} (bottom panel). 
The general trends of the temporal growth and thinning of the seasonal polar cap are plausible.
In addition, the $\scli{\mathrm{\small MAD}}{s}$ has significantly reduced from $\sim$1.5~m to just 4.9~cm. 
Noticeable are the line-shaped clusters of single measurements (black dots in bottom panel of Fig.~\ref{fig:height_change_bin_region_F}), each cluster is formed by height differences at footprints along one specific profile segment. 
The vertical length of these clusters varies between $\sim$0.1~m and $\sim$0.3~m with an average value of $\sim$0.2~m. 
As the accuracy of height differences from the local co-registration is $\sim$10~cm, the observed variation within the cluster may reflect the local variations of the CO$_2$ snow/ice cover condensation/sublimation process. 
Meanwhile, a random down-sampling of the footprints to some extent should not compromise the results but can hugely boost time-efficiency of the processing.

\begin{figure}[H]
\centering 
\includegraphics[scale=0.45]{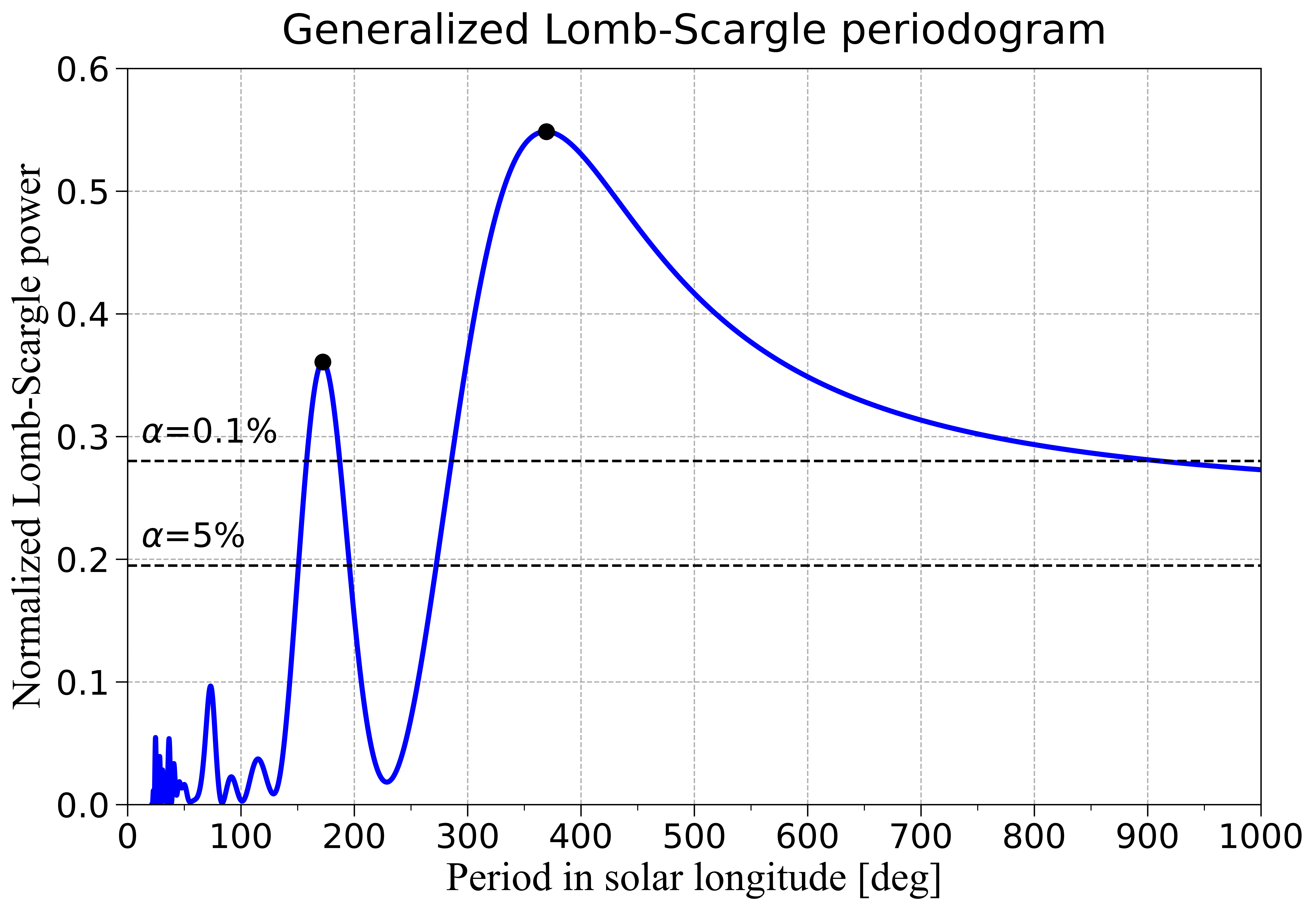}
\caption{Periodogram of the obtained height change time series in the research region from the generalized Lomb–Scargle algorithm. 
Two highest peaks are marked by black dots. 
The normalized power values corresponding to false alarm probabilities of 0.1$\%$ and 5$\%$ (horizontal dashed lines) indicate high significance of the observed peaks.}
\label{fig:periodogram}
\end{figure}

\quad Two previous results from MOLA, i.e., \cite{smith2001co2} (green line) at annulus 86.5$\degree$S and \cite{aharonson2004depth} (light blue line) at annulus 86$\degree$S, are shown in Fig.~\ref{fig:height_change_bin_region_F} for comparison. 
Although both have exclusively utilized the MOLA dataset, large discrepancies can be found between these two results. 
Our result from the "F\_LC"-method post-corrected by the bi-RPCA procedure generally resembles that of \cite{aharonson2004depth} which is consistent with contemporary measurements from epithermal neutrons and GCMs (see Fig.~8 in \cite{aharonson2004depth}). 
The peak-to-peak height variation within this research region by our estimates is $\sim$2~m, which is 0.5~m less than that of \cite{aharonson2004depth}. 
While no net height loss has been resolved of the underlying perennial polar cap, the peak height of the seasonal polar cap at the end of southern winter of MY25 is 0.18~m higher than that of MY24. 
\cite{aharonson2004depth} resolved an "off-season" accumulation from $\sclii{L}{s}=270\degree$ in the beginning of southern summer of MY24 to $\sclii{L}{s}=30\degree$ in the southern fall of MY25. 
They considered it as suspicious for a lack of viable explanations or confirmation through other independent observations. 
In our result this accumulation is less pronounced, especially in the beginning of the southern fall. 
However, an abrupt accumulation with a height increase of $\sim$0.8~m at the midst of southern summer is resolved. 
Furthermore, our result resolves a pronounced "pit" (transient height accumulation) with a magnitude of $\sim$0.5~m centered at $\sclii{L}{s}=210\degree$ of MY24. 
In contrast, a rather flat curve is resolved by \cite{aharonson2004depth}. 
This could be due to the fact that cross-overs acquired within 15 days (compare to 5 days used in this study with pseudo cross-overs) have been used in their global cross-over adjustment.
This larger interval may smooth out the delicate temporal height variation signal observed in our work.

\quad To assess the characteristics of the periodicity in the derived temporal height change, we applied the generalized Lomb–Scargle periodogram that is used to study the periodicity of unevenly sampled data \citep{vanderplas2018lomb-scargle}. 
The resultant periodogram is shown in Fig.~\ref{fig:periodogram}.
Two peaks corresponding to periods of 369.6$\degree$ (annual component) and 172.2$\degree$ (semi-annual component) in terms of solar longitude stand out, with amplitudes of the sinusoidal fittings of 0.52~m and 0.41~m, respectively. 
The peak significance as measured by the false alarm probability from bootstrap simulation for both of the aforementioned peaks are less than 0.1$\%$, strongly implying the existence of both annual and semi-annual terms in the retrieved seasonal height variation. 

\begin{figure}[H]
\centering 
\includegraphics[scale=0.6]{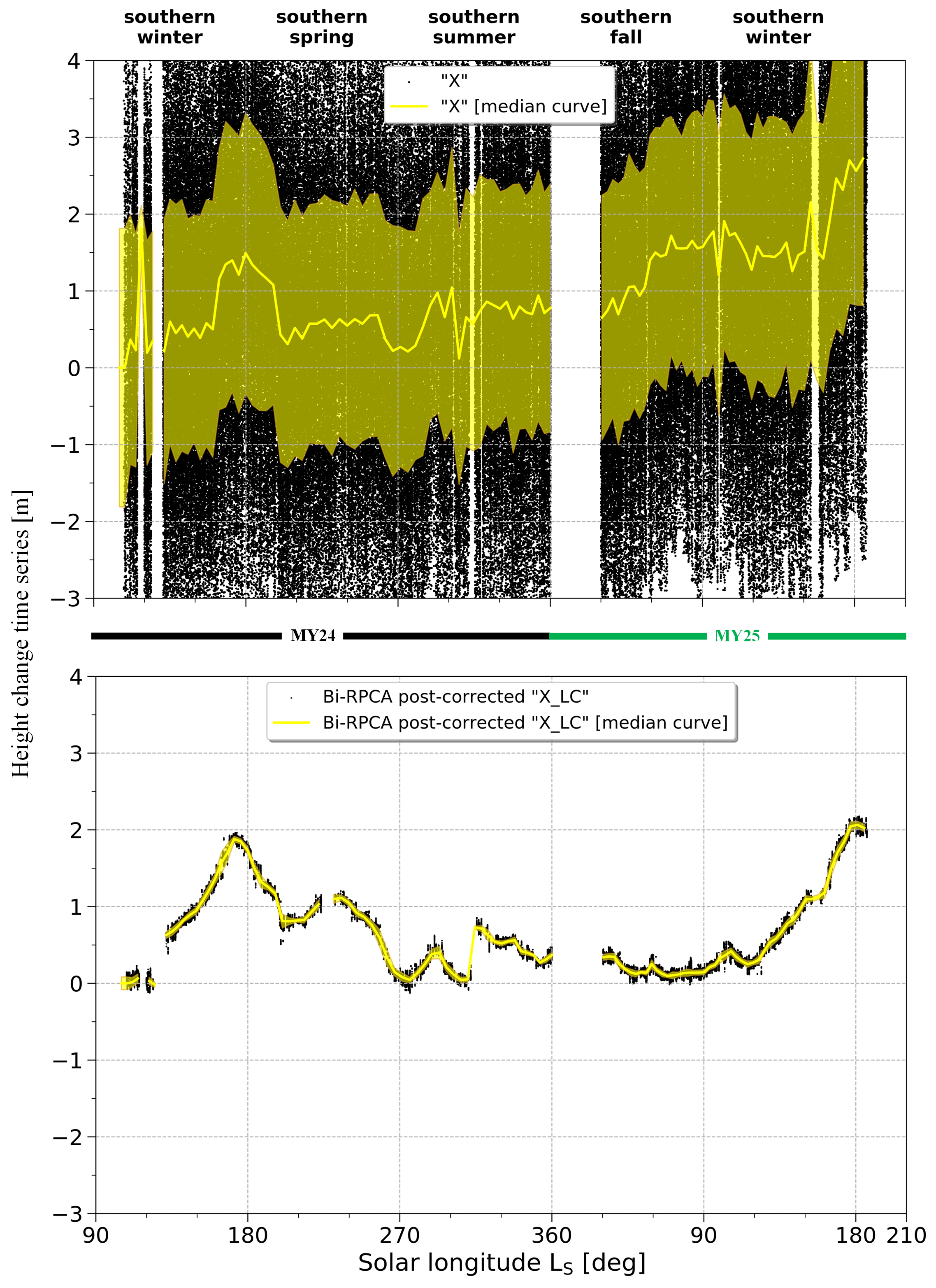}
\caption{Temporal height variation at the research region derived from "X"-method (top) and "X$\_$LC"-method post-corrected by the bi-RPCA procedure (bottom). 
Yellow shades denote the precision measured by $\scli{\mathrm{\small MAD}}{s}$.}
\label{fig:res_X}
\end{figure}

\quad Initially, there exist a sum of 258,069 cross-overs from crossings of 1,740 profiles passing through the research region (compare to 194,109 footprints). After various filtering processes in "X" and "X$\_$LC" methods, respectively (refer to Secs.~\ref{sec:co-reg_MOLA} and \ref{sec:median-binning} for the filtering processes), 229,820 and 235,986 height differences are acquired, respectively.
Comparison of the results for "X" and  "X$\_$LC"-method post-corrected by the bi-RPCA procedure, are illustrated in Fig.~\ref{fig:res_X}.
Due to the remaining errors in the reprocessed MOLA footprints and the interpolation errors due to the large distance between neighboring MOLA footprints, the mean $\scli{\mathrm{\small MAD}}{s}$ precision for "X" is at $\sim$1.5~m, a values close to that obtained by the "F"-method. 
Another considerable error source is the simple treatment in the "X"-method that assigns the observables at two different time stamps as the height differences at the cross-overs just with opposite signs. 
Similar to the "F"-method, the temporal median curve of the "X"-method is unbalanced and does not repeat at the resolved southern winters. 
After the integration of the local co-registration strategy in "X$\_$LC"-method and the bi-RPCA post-correction procedure, the noise becomes much less obvious with the $\scli{\mathrm{\small MAD}}{s}$ reduced to 4.9~cm for "X$\_$LC"-method, which is at comparable level as in the "F$\_$LC"-method. 
In addition, the temporal height variation for the "X$\_$LC"-method is balanced and nearly identical as that of the "F$\_$LC"-method.

\quad In summary, both methods, "F$\_$LC" and "X$\_$LC", are capable of generating much less noisy results than that of "F" and "X" methods, demonstrating the feasibility and benefits of the proposed local co-registration strategy and bi-RPCA post-correction procedure. 
The similarity of the results from the "F$\_$LC" and "X$\_$LC" methods is mainly due to the research region's vicinity to the MGS inclination limit. 
Profiles overwhelmingly merge and overlay at this research region, thus both footprints and cross-overs are the most densely populated with the latter even outnumbers the former (258,069 cross-overs compared to 194,109 footprints). 
Thus, the information of the MOLA measurements have been fully utilized in both of these methods and similar satisfactory results can be expected. 
In fact, the introduction of the reference DTM and the local co-registration strategy has reconciled these two methods to one that spatially samples temporal height differences of the local profile segments with respect to the reference DTM. 
The differences are just the locations of samplings, i.e., at either footprints or cross-overs, and the frequency (or number) of these samplings. 
As long as statistical independent samplings have been made (at least one along each individual profile segment), all of the information from the laser altimetry can be exploited. 
Excessive redundant samplings can not further enhance the results. Downsampling, to some extent, can be performed in order to reduce the computation load. 
Indeed, the extended length of the local profile segments ($\sim$600 footprints) means that height difference measurements at footprints along one specific local profile are highly inter-dependent. 
Thereby, the quality of the derived temporal height variation is largely determined by the number of individual profiles available and by the accuracy of these profiles. 
In addition, as the number of pseudo cross-overs increases quadratically with the number of profiles.
An increased number of individual profiles would be of large benefit for the RPCA post-correction process, further improving the precision of the obtained temporal height variation.

\quad However, in regions towards the equator, where tracks are nearly meridional and do not necessarily cross each other, the "X$\_$LC"-method may even fail to yield satisfactory results, and the "F$\_$LC"-method can be extremely useful. 
Another strength of the "F$\_$LC"-method compared to "X$\_$LC"-method is that the tedious and time-consuming cross-over searching process of the latter can be well spared. 
Overall, based on the results we strongly recommend "F$\_$LC"-method over the "X$\_$LC"-method in terms of mapping the seasonal height variations of Martian poles.

\section{Discussion}
\label{sec:disc}

\subsection{Number of footprints required to stabilize the co-registration}
\label{sec:min_points_stability}

\begin{figure}[H]
\centering
\includegraphics[scale=0.6]{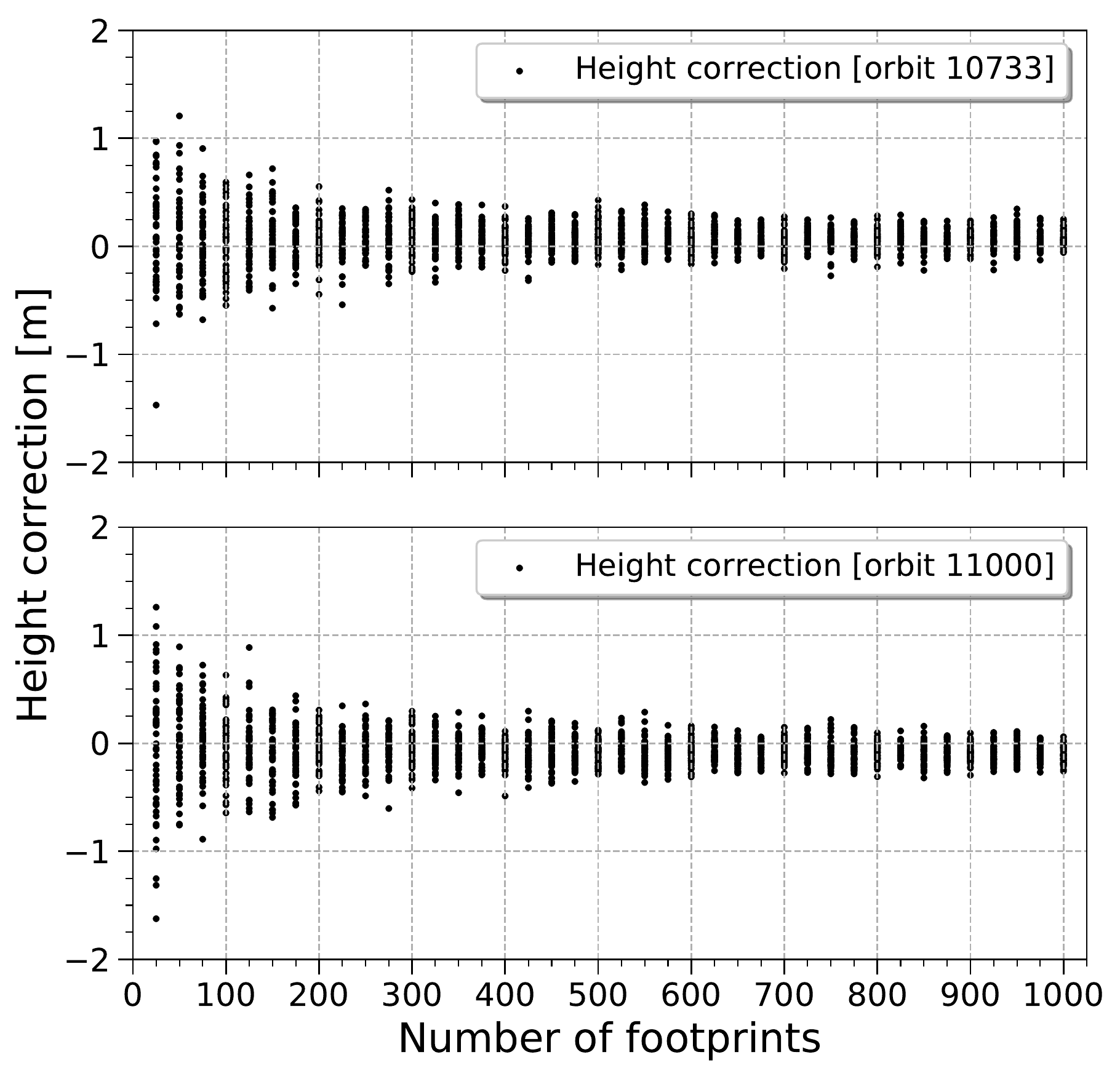}
\caption{Height corrections plotted against the number of footprints in the sub-samples randomly selected from two typical nadir-pointing profiles as 10,733 (top) and 11,000 (bottom). The convergences are observed when the number of footprints amounts to $\sim$400 to 600.}
\label{fig:min_points_stability}
\end{figure}

\quad In this research, a total of 600 footprints are chosen to form local profile segments and to use them in the local co-registration processes. 
Due to the circular orbit of MGS and constant MOLA firing rate (10~Hz), in practice, this is done by including footprints acquired within $\pm$30 seconds of the footprint or cross-over in question. 
For nadir-pointing profiles this is equivalent to about 90~km or about 300 footprints on each side. 
These numbers are chosen after an analysis of the minimum number of footprints required to stabilize the co-registration. 
In this analysis we first select 1,500 consecutive footprints along a profile to form a complete sample. 
Then, we generate sub-samples by randomly selecting MOLA footprints from the complete sample. 
We study sub-samples sizes from 25 to 1,000 with a step size of 25. 
Subsequently, these sub-samples are co-registered to the self-registered MOLA reference DTM. 
To account for biases due to the random selection, this whole process is repeated for 50 times. 
Fig.~\ref{fig:min_points_stability} shows the height corrections obtained from the co-registrations against the number of footprints in the sub-samples for two typical nadir profiles (10,733 and 11,000). 
For both profiles the parameters begin to converge when the number of footprints in the sub-sample ranges from $\sim$400 to 600. 
It is worthy to note that the number of consecutive footprints in the complete sample does not affect the general trend of the height corrections and therefore the convergence pattern.

\quad This relatively large number of footprints required for accurate co-registration implies a coarse "effective spatial resolution" of the derived temporal height variations and is a major limitation of the proposed local co-registration strategy. 
This issue could be partially overcome by incorporating stereoclinometric DTMs with higher spatial resolution and better accuracy (see alo Sec.~\ref{sec:CTX_HiRISE_DTMs}). 
In addition, future high frequency laser altimeter missions to Mars can be expected to overcome this issue. 
A good example on Earth is the Advanced Topographic Laser Altimeter System (ATLAS) onboard ICESat-2. 
With a footprint interval of less than 1~m, it can enable a stable co-registration on the scale of just several kilometers \citep{nan2019icesat2}.

\subsection{Quantification of the errors induced by interpolation and profile lateral shifts}
\label{sec:quantification_errors}

\begin{figure}[H]
\centering 
\includegraphics[scale=0.4]{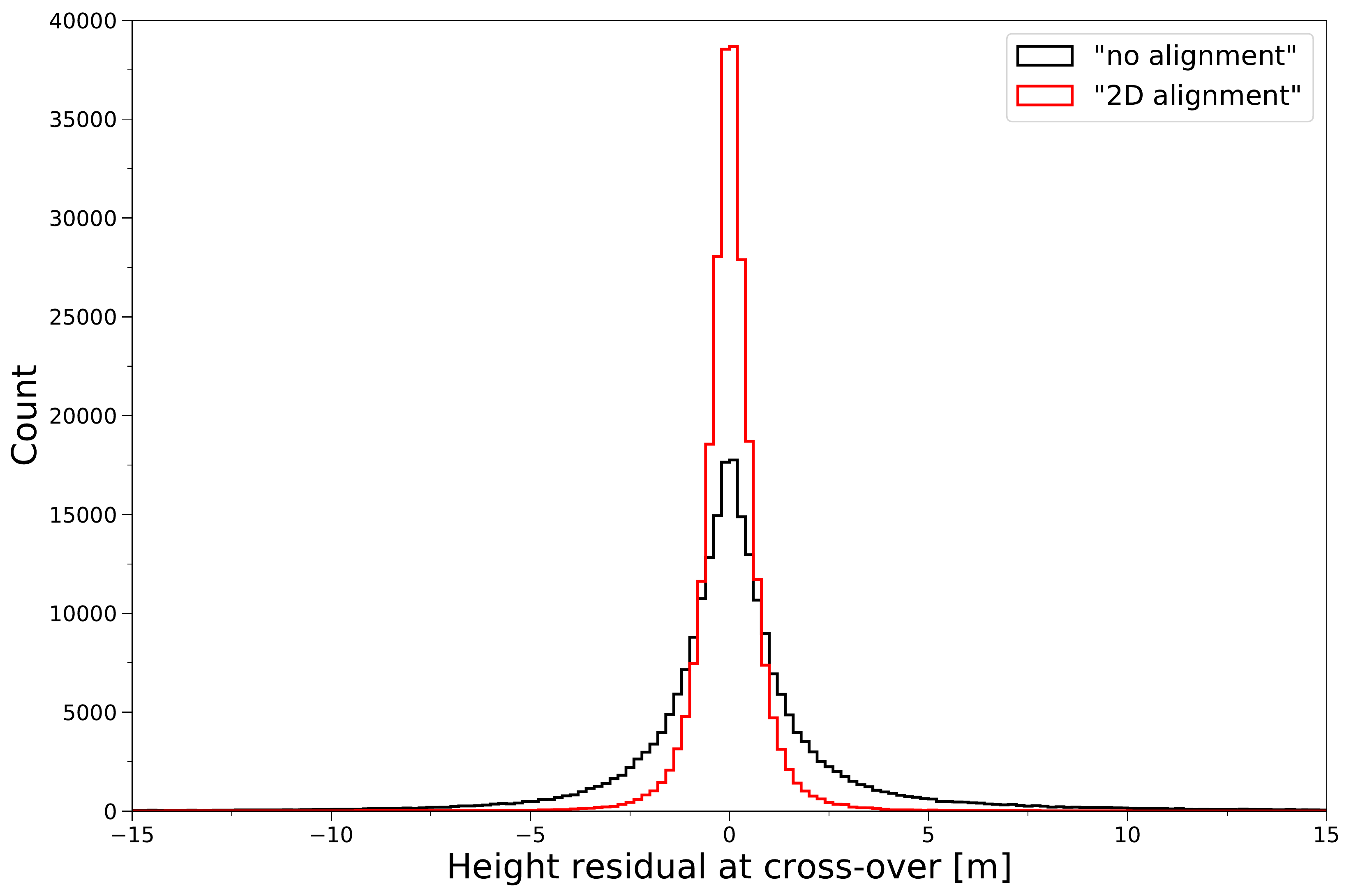}
\caption{Histogram of the cross-over height residuals with "no alignment" and "2D alignment". Note that the RMS height residual with "3D alignment" is almost zero, thus the corresponding histogram is not shown.}
\label{fig:error_quantification}
\end{figure}

\quad To quantify the impacts of interpolation and lateral shifts of the MOLA profiles on the temporal height differences at cross-overs, we carry out a simulation experiment for profile segments falling within the research region of [85.75$\degree$S, 86.25$\degree$S, 300$\degree$E, 330$\degree$E]. 
We first create synthetic MOLA profiles, where the heights at the footprints are extracted from the self-registered MOLA reference DTM after co-registering the profiles to it (see Sec.~\ref{sec:self_registration_MOLA}). 
However, the coordinates of the footprints are taken from the original profiles before the co-registration, preserving lateral shifts with respect to the reference DTM. 
Thus, the theoretical height differences at the cross-overs should vanish, and any deviations should be interpreted as height residuals due to pure errors, e.g., interpolation errors or lateral shifts of the profiles. 
The height residuals at the cross-overs of the synthetic profiles are obtained in three ways: (1) "no alignment": linear interpolation of the heights at the cross-overs from neighboring footprints along the ascending and descending pairs, and subtracting one from another to retrieve the height residuals at the cross-overs.
(2) "2D alignment": laterally co-registering the laser profiles to the reference DTM to account for lateral shifts of the profiles and then using "no alignment" to resolve the height residuals at the cross-overs.
(3) "3D alignment": co-registering the laser profiles to the reference DTM in both lateral and radial directions and assigning the height residuals at each of the cross-overs as the differences in radial corrections for the intersecting profile pairs. 
The lateral corrections of each of the profiles in "2D alignment" and "3D alignment" are identical. 
Histograms of the height residuals at cross-overs for the different treatments are shown in Fig.~ \ref{fig:error_quantification}. 
The RMS height residual is 3.74~m with "no alignment", 1.40~m with "2D alignment", and 0~m with "3D alignment". 

\quad In the context of this simulation, the obtained RMS value of $\sim$3.74~m represents the combined impacts of interpolation errors and profile lateral shifts, while $\sim$1.40~m stands only for the errors exerted by the interpolation between footprints. 
Note that the magnitudes of the interpolation errors are positively related to the surface slope and roughness of a specific area, which can be much larger for regions rich in craters and escarpments. 
If treating these two errors sources as uncorrelated, the profile lateral shifts (in fact a mean magnitude of 49~m in line direction and 38~m in sample direction, see Fig.~\ref{fig:line_sample_height_correction}) can lead to a RMS height residual on cross-overs of $\sim$3.47~m, which is roughly two times that of the interpolation errors in this region. 
The interpolation distances to the cross-overs from the four neighboring footprints are directly affected by the profile lateral shifts. In particular, for cross-overs with shallow intersection angles, even a limited component of the lateral shifts perpendicular to the tracks can lead to a large displacements of the cross-over locations. 
Hence, the height residuals at cross-overs can be quite sensitive to lateral geolocation errors of the profiles in question. 
This highlights the necessity of addressing these profile lateral shifts with the proposed local co-registration strategy.

\subsection{Systematic temporal bias}
\label{sec:sys_bias_appex}

\begin{figure}[H]
\centering 
\includegraphics[scale=0.8]{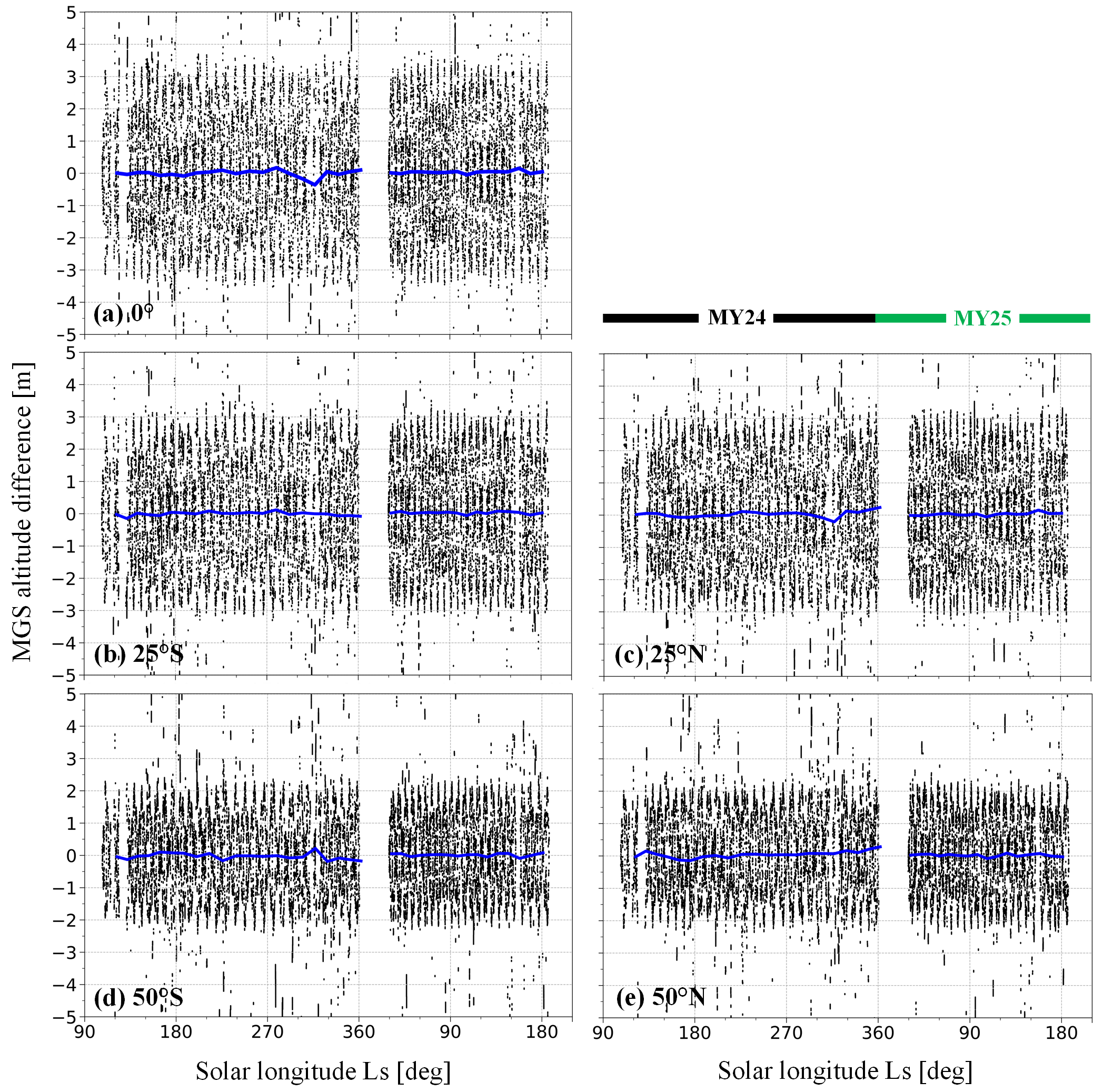}
\caption{Comparison of MGS altitude of the old orbit documented in MOLA PEDR and the new orbit adopted for the reprocessing of MOLA dataset at different annuli centered at (a) 0$\degree$, (b) 25$\degree$S, (c) 25$\degree$N, (d) 50$\degree$S, and (e) 50$\degree$N. Each black dot denotes a MGS altitude difference at the acquisition time of a specific footprint while blue lines represent the median-binned temporal curves.}
\label{fig:orbit_comparison}
\end{figure}

\begin{figure}[H]
\centering 
\includegraphics[scale=0.8]{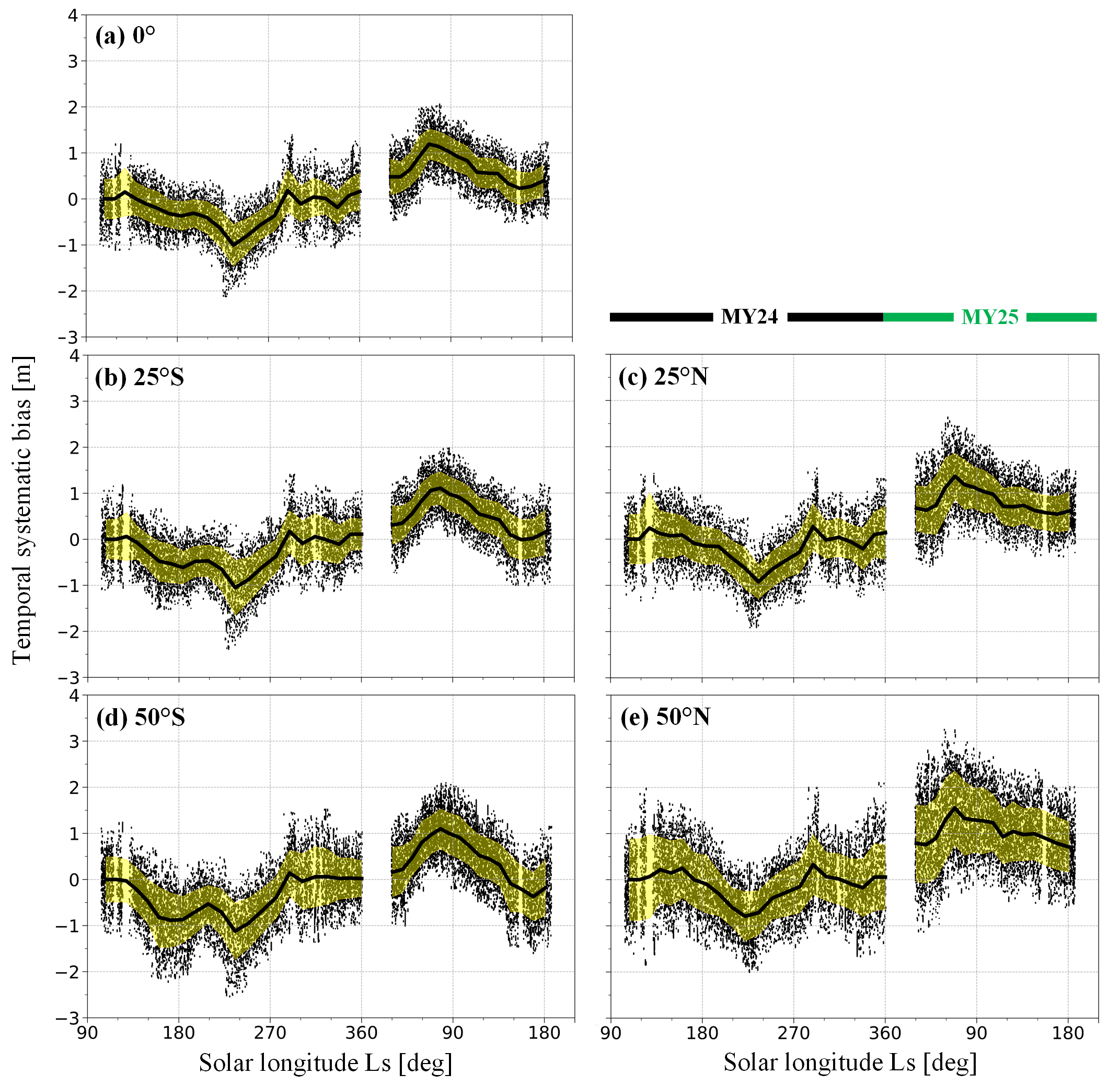}
\caption{Systematic temporal bias in MOLA footprint heights by median-binning the radial corrections from the global cross-over analysis from \cite{neumann2001cross-over} at different annuli centered at (a) 0$\degree$, (b) 25$\degree$S, (c) 25$\degree$N, (d) 50$\degree$S, and (e) 50$\degree$N. Black lines represent the median-binned temporal curves while the yellowish shades denote the precisions measured by $\scli{\mathrm{\small MAD}}{s}$.}
\label{fig:sys_bias}
\end{figure}

\quad Temporal evolution of the MGS altitude differences between old (the one documented in MOLA PEDR) and new orbit models \citep{konopliv2006orbit} in annuli 0$\degree$, 25$\degree$S/N and 50$\degree$S/N, where the effects of seasonal CO$_2$ condensation/sublimation are negligible, is shown in Fig.~\ref{fig:orbit_comparison}. 
We have not observed any obvious temporal trend (some relatively large deviations from zero are mainly due to data gaps or being near the solar conjunction).
This means that the systematic temporal bias in MOLA footprint heights reported by \cite{smith2001co2} remains even after the reprocessing of the MOLA dataset in this study. 
Initially, we can obtain the systematic temporal bias by median-binning temporal radial corrections from the global cross-over adjustment that are documented in MOLA PEDR \citep{neumann2001cross-over}. 
The results for the aforementioned equatorial and tropical annuli are shown in Fig.~\ref{fig:sys_bias}. 
The precisions represented by $\scli{\mathrm{\small MAD}}{s}$ range from 40 to 100~cm. 
Although the systematic temporal bias exhibits similar general trends in these annuli within 50$\degree$S/N, significant local variations do exist.

\begin{figure}[H]
\centering 
\includegraphics[scale=1]{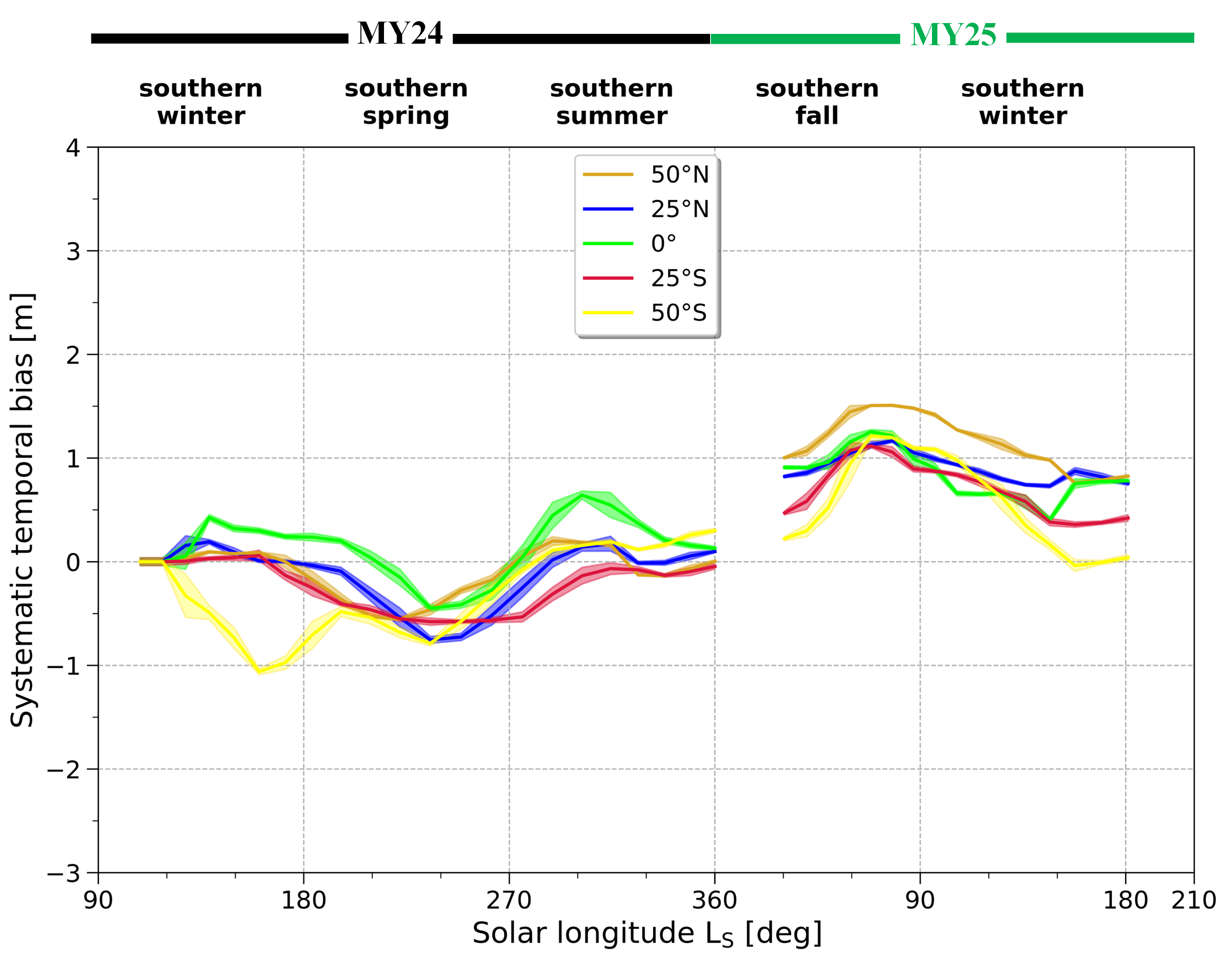}
\caption{Comparison of systematic temporal biases in MOLA footprint heights from RPCAs performed at annuli centered at 0$\degree$, 25$\degree$S/N, and 50$\degree$S/N. Trends are represented by median-binned temporal curves and the shaded denote the precisions measured by $\scli{\mathrm{\small MAD}}{s}$.}
\label{fig:sys_bias_variation}
\end{figure}

\quad Quoted precision of the retrieved CO$_2$ seasonal height variation is simply based on statistics and do not include allowance for unaccounted systematic errors. 
For the bi-RPCA procedure, it is assumed that the temporal systematic bias is constant at the polar regions and equals that obtained at annulus 50$\degree$S.
To assess the validity of this assumption, we also taken advantage of the proposed RPCA procedure as in Sec.~\ref{sec:RPCA_50S}, and apply it to annuli 0$\degree$, 25$\degree$S/N, and 50$\degree$S/N to unambiguously determine the systematic temporal biases. 
As in annulus 50$\degree$S, an optimal regularization strength of 1 is introduced to stabilize the inversions. 
Results are shown in Fig.~\ref{fig:sys_bias_variation}. The $\scli{\mathrm{\small MAD}}{s}$ precisions of these obtained biases are better than 10~cm which are much better than that obtained from the temporal radial corrections in the global cross-over adjustment \citep[Fig.~\ref{fig:sys_bias};][]{neumann2001cross-over}. 
While the general trends are similar, variations up to 1.5~m (at $\sclii{L}{s}=160\degree$ in MY24) can be observed. 
Thus, the systematic temporal bias do vary significantly with latitudes and the aforementioned assumption does not stand to some extent. 
Judging from the evolution of the bias at these sampled latitudes, there exists the possibility that our result has underestimated the seasonal height variation magnitude at the polar regions. 
Thus, locating the root cause for this systematic bias and amply compensate for it hold as the key to unbiased CO$_2$ seasonal height variation measurements. 
We defer this to future work.

\subsection{Potential of using CTX, HiRISE DTMs and SHARAD radar altimetry}
\label{sec:CTX_HiRISE_DTMs}
\quad In \ref{sec:height_change_HRSC_appex}, we also demonstrate the feasibility of using the High-Resolution Stereo Camera (HRSC) DTMs \citep{putri2019new} instead of the self-registered MOLA reference DTM as the static mean reference surface in mapping seasonal height variations. 
However, the achieved precision of 17~cm is much worse than 4.9~cm when using the self-registered MOLA reference DTM.
The HRSC DTMs feature a grid resolution of 50~m and a possibly significantly lower effective resolution. 
The nominal effective footprint size of MOLA has been stated as 150~m which corresponds to a divergence angle of 93~µrad. 
However, two post-launch studies suggested lower divergence angles due to a hot spot in the laser beam, resulting in footprint sizes of 75~m and 50~m, respectively \citep{neumann2003mola, kim2008roughness}. 
Thus, Context Camera \citep[CTX;][]{malin2007ctx} and High Resolution Imaging Science Experiment \citep[HiRISE;][]{mcewen2007HiRISE} DTMs with higher spatial resolution can be expected to be beneficial in the local co-registration processes and hence the height variation measurements. 
Hence, less DTM-induced errors and less number of consecutive MOLA footprints needed to stabilize the co-registration imply a higher spatial effective resolution of the derived seasonal height variation maps. 
As availability and spatial coverage of CTX and HiRISE stereo images and DTMs is continuously increasing, the additional benefits of their use should be investigated in future work \citep{tao2018DTM}.

\quad It is not possible to derive a long-term height change time series from MOLA data due to its operation time being limited to $\sim2$ Earth years. 
However, this can be achieved by using data collected by the SHAllow RADar \citep[SHARAD;][]{seu2007sharad} mounted on the Mars Reconnaissance Orbiter (MRO). 
Combined, MOLA and SHARAD represent an survey of Mars' polar regions over a period of more than two decades (although with gaps from June 2001 to December 2006). 
Although designed primarily as a subsurface sounder, the 20~MHz center frequency and 10 MHz bandwidth of SHARAD allows its use as a radar altimeter \citep{mouginot2009marsis, steinbrugge2019sharad}. 
After post-processing involving range compression, ionospheric correction, and “unfocused SAR processing”, the altimetric ranges can be determined by basically comparing the radargrams to the simulated cluttergrams using the self-registered MOLA reference DTM. 
With a firing frequency of 700 Hz and 3600 samples per shot, we can expect a high volume of available altimetric measurements. 
Despite the coarse inherent range resolution of the instrument (15 m), spatial and temporal averaging combined with the proposed local co-registration strategy and the RPCA post-correction procedure can be expected to substantially improve the precision of the obtained seasonal height variations to sub-meter level. 
Once proven feasible, the long-term evolution of the volatile cycle and stability of the residual polar caps can be critically analyzed.

\subsection{Potential applications}
\quad Apart from measuring the height variations of the Martian poles, the local co-registration strategy and RPCA post-correction procedure employed in this study can also be readily adopted for estimating the height change of the ice sheets in terrestrial cryosphere and thereby their mass balance. 
For this application, commercial high-resolution reference DTMs created from stereophotoclinometry using satellite imagery with sub-meter resolution are already available \citep{howat2019}. 
Particularly, by introducing in high-resolution DTMs, the local co-registration approach is immune to the side impact of significant gaps between repeat laser tracks (e.g., up to hundreds of meters for ICESat), and the modelling error by approximating some of the rough terrain with low-order polynomials from the sparse laser tracks \citep{schenk2012}. 
This can be proven to be especially useful in the marginal regions of the ice sheets where the height change signal is the strongest, but traditional approaches using laser altimetry and radar altimetry prove less effective due to significant slope and roughness \citep{pritchard2009, helm2014, schroder2019}.

\quad Besides the height change, another major application of the proposed methods are the measurement of the tidal Love number $h_2$ of various planetary bodies, e.g., the Moon, Mercury, and Ganymede. 
Firmly constrained $h_2$ values can facilitate tapping into the interior structures of the target bodies, for example to confirm or deny the existence of a subsurface ocean at Ganymede \citep{steinbrugge2015tide} or to constrain the interior structure and rheology Mercury's core \citep{thor2020bela, steinbruegge2021}.
Currently, the measurements of body tides predominantly rely on temporal height differences at cross-overs \citep[e.g.,][]{mazarico2014tide, steinbrugge2015tide, steinbrugge2018europa, bertone2021}. 
However, for slowly rotating bodies like Moon and Mercury, the polar orbit of the spacecraft implies that much less cross-overs are available in the equatorial and tropical regions where significant tidal amplitudes are expected \citep{mazarico2014tide, steinbrugge2018bela}. 
Furthermore, the intersection angles of profiles forming these non-polar cross-overs can be shallow, which makes them vulnerable to cross-track components of profile lateral shifts due to residual errors. 
Even limited cross-track displacements are capable of shifting these intersection points dramatically upwards or downwards along the near-parallel profiles, compromising the measured cross-over height differences \citep[][see also Sec. \ref{sec:quantification_errors}]{xiao2020geolocation}. 
The proposed approach of the "F\_LC"-method performing at individual footprints in combination with the RPCA post-correction procedure can contribute to alleviating these issues in these equatorial and tropical regions.

\quad Recently, the direct differencing between a laser altimeter footprint and the underlying DTM has been proposed and validated in the refinement of Kaguya's orbit \citep{goossens2020orbit}. 
While these so called direct altimetry measurements do not suffer from the interpolation errors like at the cross-overs, it is subject to the interpolation errors of some interpolated DTM pixels and the geolocation error of each single laser footprint. 
Thus, the proposed local co-registration strategy can be of use in the context of the direct altimetry to refine spacecraft orbits. 
Another important application can be merging of laser altimeter profiles and DTMs from stereoscopic pairs to obtain refined topographic models with greater details \citep{barker2016lola}.

\section{Conclusions}
\label{sec:conc}
\quad In this research, we have investigated both the feasibility and benefits of retrieving seasonal CO$_2$ snow/ice height variations in Mars's polar areas using MOLA data.
Thereby, we implemented a local co-registration strategy of MOLA laser altimeter profiles to a reference DTM produced by self-registered MOLA profiles.
First, we reprocessed the MOLA profiles using updated MGS orbit and Mars rotational models. 
Then, we self-registered these reprocessed MOLA profiles to form a coherent reference DTM for the Martian South Pole poleward of 78$\degree$S. 
Subsequently, individual reprocessed MOLA profile segments centered at either footprints or cross-overs were co-registered to the self-registered MOLA reference DTM. 
The temporal height differences at either footprints or cross-overs are then assigned as the height corrections from these local co-registration processes. 
In addition, the bi-RPCA post-correction procedure was proposed and applied to remove both a temporal systematic bias and other residual errors in the reprocessed MOLA profiles and the self-registered MOLA reference DTM. 
Finally, seasonal height variation were obtained by median-binning the temporal height differences. 
The proposed co-registration can largely avoid the interpolation errors, which can result in a RMS height residual of $\sim$1.40~m at cross-overs in the research region. 
In addition, it can naturally correct for residual lateral shifts of the laser profiles, which can lead to RMS height residual of $\sim$3.47~m at cross-overs in the research region.

\quad The precision of the derived height change time series in the research region on the residual polar cap has been improved from $\sim$1.5~m to $\sim$4.9~cm after the adoption of the local co-registration strategy and the bi-RPCA post-correction procedure, an enhancement by a factor of $\sim$30.
The maximum height variation is estimated at $\sim$2~m. 
The off-season accumulation of snow and ice in the southern summer pointed out by \cite{smith2001co2} and \cite{aharonson2004depth} is also resolved, but it is less evident than in the aforementioned studies. 
However, a pronounced "pit" (transient height accumulation) of $\sim$0.5~m in magnitude centered at $\sclii{L}{s}=210\degree$ in southern spring at MY24 is observed. 
It is expected that seasonal CO$_2$ snow/ice height variations with high spatial and temporal resolution at the entire Martian Poles can be unambiguously mapped using the proposed local co-registration strategy and post-correction procedure.

\quad In future work, the potential of high-resolution CTX and HiRISE DTMs in the seasonal height variation mapping should be assessed. 
In addition, the possibility of deriving long-term height variations using SHARAD radar altimetry shall be explored. 
A wide range of other applications can be envisioned to benefit from the proposed local co-registration strategy complemented by the RPCA post-correction procedure, e.g., measuring tidal deformation for various planetary bodies, constraining elevation change and mass balance of the terrestrial cryosphere, merging of different topographic data sets as well as refining spacecraft orbits.

\section*{Acknowledgements} 
\label{Acknowledgements}
\quad The authors would like to thank Gregory Neumann (NASA GSFC) and Philipp Gl{\"a}ser (TU-Berlin) for helpful discussions. 
H. Xiao thanks the China Scholarship Council (CSC) for financial support of his Ph.D. study in Germany. 
Part of this research was carried out within the framework of the DLR BigData Querschnittsplattform. 
F. Schmidt acknowledges support from the ``Institut National des Sciences de l'Univers'' (INSU), the ``Centre National de la Recherche Scientifique'' (CNRS) and ``Centre National d'Etudes Spatiales'' (CNES) through the ``Programme National de Plan{\'e}tologie''. 
We also acknowledge the MOLA and HRSC teams for providing the great products that enabled this research.

\section*{Author Contributions}
\quad AS and HX conceptualized the research; HX and AS contributed to methodology; AS and JO supervised the study; HX, AS and GS developed the software; HX and AS performed the validation and performance analysis; HX prepared the initial draft; AS, GS, RT, FS and JO contributed to review and data analysis; HX and JO acquired funding.

\section*{Data Availability}
 \quad The MOLA PEDR dataset (Version L) is available from the Planetary Data System (PDS) Geosciences Node at https://pds-geosciences.wustl.edu. The MGS trajectory and attitude kernels can be obtained from NASA NAIF at https://naif.jpl.nasa.gov/naif. The HRSC DTMs utilized in the Supplement are available for downloading at ESA Planetary Science Archive (PSA) at https://www.cosmos.esa.int/web/psa/UCL-MSSL$\_$iMars$\_$HRSC$\_$v1.0.

\section*{ Conflict of Interest}
\quad The authors declare that they have no conflict of interest.

\bibliography{references}

\newpage\null\thispagestyle{empty}\newpage

\section*{Supplementary Materials}

\beginsupplement

\setcounter{table}{0}  
\setcounter{figure}{0}
\setcounter{equation}{0}
\setcounter{section}{0}

\renewcommand{\thetable}{S\arabic{table}}
\renewcommand{\thefigure}{S\arabic{figure}}
\renewcommand{\theequation}{S.\arabic{equation}}
\renewcommand{\thesection}{Supplement \Alph{section}}

\section{Co-registration and self-registration techniques}
\label{sec:co-regist_self_regist}

\quad In this section, we first introduce the co-registration technique used to align laser profiles to a reference Digital Terrain Model (DTM), then building upon the co-registration technique the self-registration of the laser profiles themselves to a coherent reference DTM is presented. 

\subsection{Co-registration of laser profiles}
\label{sec:co-reg_tech}
\quad The co-registration technique relates the individual time-dependent, discrete laser profiles to the static, continuous surface modelled by the DTM. 
By doing so, offsets of the laser profiles mainly induced by residual orbit, pointing, and timing errors can be corrected and the Root Mean Square (RMS) height residuals between the two datasets at the laser footprints' locations can be minimized:
\begin{linenomath*}
\begin{equation}
\arg \min_{\hat{\mathbf{p}}} \,\sqrt{\dfrac{\sum_{i=1}^{n}\left(h^{i}_{\mathrm{DTM}}(\mathbf{p})-h^{i}_{\mathrm{LA}}(\mathbf{p})\right)^{2}}{n}} \,,
\end{equation}
\end{linenomath*}
where $h^{i}_{\mathrm{LA}}$ refer to the height of the $i^{\mathrm{th}}$ laser altimeter footprint and $h^{i}_{\mathrm{DTM}}$ is the interpolated DTM height at the location of this footprint, $n$ denotes the number of footprints along the profile, and $\mathbf{p}$ represents the tangible set of parameters to be solved, e.g., map coordinates of the DTM, rotational angles, and laser altimeter pointing angles \citep{glaeser2013, stark2015feasibility, tang2019zy3}. 
As this process is mostly nonlinear, i.e., the derivatives of the functional model to be minimized with respect to the model parameters to be solved are not constant, an iterative least squares approach is employed to solve this inverse, non-linear problem \citep{tarantola1982leastsquares}:
\begin{linenomath*}
\begin{equation}
    \hat{\mathbf{p}}_{k+1}=\hat{\mathbf{p}}_{k}-\left(\mathbf{G}_{k}^{\mathrm{T}}\mathbf{C}_{g}^{-1} \mathbf{G}_{k}\right)^{-1} \mathbf{G}_{k}^{\mathrm{T}} \mathbf{C}_{g}^{-1} \mathbf{g}\left(\hat{\mathbf{p}}_{k}\right)\,,
\label{eq:iter_solution}
\end{equation}
\end{linenomath*}
where $k$ represents the $k^{\mathrm{th}}$ iteration, $\mathbf{C}_{g}$ is the covariance matrix of the observations, $\mathbf{G}_{k}$ is the partial derivatives matrix introduced to linearize the functional model and $\mathbf{g}\left(\hat{\mathbf{p}}_{k}\right)$ denotes the vector of adjusted height residual observables to be minimized in the functional model. 
In this study, we make use of two different parameterizations, which are introduced in the following sections: a parameterization in map coordinates, and a parameterization in map coordinates and a height trend.

\subsubsection{Co-registration parameterized in map coordinates}
\label{sec:local_co-registration_in_map_coords}

\quad The co-registration parameters can be defined in map coordinates of the DTM, i.e. line $\delta l$, sample $\delta s$, and height $\delta h$ \citep{glaeser2013}
\begin{linenomath*}
\begin{equation}
\begin{split}
\mathbf{p}=\begin{bmatrix}
\delta l,\,
\delta s,\,
\delta h
\end{bmatrix}^{\rm{T}}\,,
\end{split}
\label{eq:co_regi_params}
\end{equation}
\end{linenomath*}
which is straightforward and can be easily implemented. 
The functional model at the location of the $i^{\mathrm{th}}$ footprint along the profile has the form
\begin{linenomath*}
\begin{equation}
\mathbf{g}^{i}(\mathbf{p})
=
h_{\mathrm{DTM}}\left(l^{i}_\mathrm{LA}+\delta l, \, s^{i}_\mathrm{LA}+\delta s\right)-\left(h^{i}_{\mathrm{LA}}+\delta h\right) \,,
\end{equation}
\end{linenomath*}
where $l^{i}_\mathrm{LA}$ and $s^{i}_\mathrm{LA}$ denote the non-integer line and sample of the $i^{\mathrm{th}}$ laser altimeter footprint projected into DTM image coordinate system. The transformation of spherical coordinates used for the laser altimeter footprint locations to image coordinates is determined by the employed map projection, e.g., a stereographic projection at the poles. The partial derivatives matrix $\mathbf{G}_{k}$ evaluated at the $k^{\mathrm{th}}$ iteration can be written as

\begin{linenomath*}
\begin{equation}
\mathbf{G}_{k}=\frac{\partial \mathbf{g}\left(\hat{\mathbf{p}}_{k}\right)}{\partial \hat{\mathbf{p}}_{k}}
=\left(\begin{array}{ccc}
\left.\dfrac{\partial h_{\mathrm{DTM}}}{\partial l}\right|_{l=l^{1}_{\mathrm{LA}_k}} & \left.\dfrac{\partial h_{\mathrm{DTM}}}{\partial s}\right|_{s=s^{1}_{\mathrm{LA}_k}} & -1 \\
\vdots & \vdots & \vdots \\
\left.\dfrac{\partial h_{\mathrm{DTM}}}{\partial l}\right|_{l=^{n}_{\mathrm{LA}_k}} & \left.\dfrac{\partial h_{\mathrm{DTM}}}{\partial s}\right|_{s=s^{n}_{\mathrm{LA}_k}} & -1 \\
\end{array}\right)\,.
\label{eq:C_matrix_map_cor}
\end{equation}
\end{linenomath*}
The accuracy of the retrieved parameters can be obtained through covariance analysis of the least squares process.

\subsubsection{Co-registration parameterized in map coordinates and height trend}
\label{sec:local_co-registration_in_map_coords_and_trend}

\quad When there is a linear temporal trend observed between the two datasets in terms of height difference, the temporal height change rate $\dot h$ can also be included as an additional model parameter. Then, the unknown parameters become
\begin{linenomath*}
\begin{equation}
\begin{split}
\mathbf{p}=\begin{bmatrix}
\delta l,\,
\delta s,\,
\delta h,\,
\dot h
\end{bmatrix}^{\rm{T}}
\end{split}
\label{eq:regi_params_lshtrend}
\end{equation}
\end{linenomath*}
and the functional model then has an additional term related to $\dot h$
\begin{linenomath*}
\begin{equation}
\mathbf{g}^{i}(\mathbf{p})
=
h_{\mathrm{DTM}}\left(l^{i}_\mathrm{LA}+\delta l, \, s^{i}_\mathrm{LA}+\delta s\right) -\left(h^{i}_{\mathrm{LA}}+\delta h + \dot h(t^{i}-t_{0})\right) \,,
\end{equation}
\end{linenomath*}
where $t^{i}$ is the ephemeris time of the $i^{\mathrm{th}}$ footprint along the profile, and $t_{0}$ is the reference epoch, e.g., the epoch of the first footprint of the profile.

\subsection{Self-registration of laser profiles}
\label{sec:self_regist}
\quad The self-registration process refines the dataset through iterations. 
For each iteration, a random subset of laser profiles is selected and then co-registered to an intermediate DTM constructed from the remaining profiles. 
Profiles with statistically gross RMS height residual will be considered as outliers and excluded from the analysis. 
After a fixed number of iterations the profiles are removed from offsets induced by orbit, pointing, and timing errors and positioned in agreement with each other. 
Finally, a self-coherent DTM can be built from these self-registered profiles and used as baseline surface for further applications. 
The iterative self-registration technique is similar to the commonly-used global cross-over adjustment, where cross-over height misfits of intersecting profiles are used as observables \citep{neumann2001cross-over}. However, typical issues present in the global cross-over analysis are resolved in the self-registration approach. 
For instance, the interpolation errors caused by large separation of laser footprints, and unfavourable geometry of highly acute intersection angles for laser altimeters onboard spacecrafts in near-polar orbit around slow-rotating celestial bodies. 
Indeed, while the cross-over observables are interpolated from their four adjacent footprints, all footprints along a profile are used to be aligned to footprints of all other profiles in its vicinity in the self-registration approach, involving more information.

\quad This technique has already been successfully applied in planetary science. 
Due to limited coverage of high-quality DTMs from stereo images, \cite{stark2018self-reg2, stark2018self-reg} have explored the self-registration of the laser profiles themselves to construct an internally consistent DTM for the estimation of rotational parameters of Moon and Mercury. 
\cite{barker2020improved} has self-registered the Lunar Orbiter Laser Altimeter (LOLA) profiles to self-consistent DTMs and examined the associated error quantification for the purpose of illumination assessment and landing site selection at Moon's South Pole.

\section{Internal consistency of the reprocessed MOLA PEDR}
\label{sec:cross-over_misfits_reprocessed}
\quad To investigate if the adoption of the recent Mars Global Surveyor (MGS) orbit model from \cite{konopliv2006orbit} and Mars rotational model from \cite{archinal2018report} could actually bring improvements to the Mars Orbiter Laser Altimeter (MOLA) Precision Experiment Data Records (PEDR) dataset, we compare the height discrepancies at cross-overs for MOLA PEDR and the reprocessed MOLA PEDR in the region of [60$\degree$S,$\,$60$\degree$N,$\,$180$\degree$E,$\,$220$\degree$E] which entails all range of latitudes between 60$\degree$S/N where limited seasonal CO$_2$ height variation is expected. To boost calculation efficiency, we concentrate on profiles acquired from March 1999 to November 1999. Before statistical analysis, these height discrepancies are edited through a 5-$\sigma$ iterative filtering process. As a result, 2.5$\%$ and 3.2$\%$ of these height discrepancies are excluded for MOLA PEDR (291,672 remain) and the reprocessed MOLA PEDR (263,448 remain), respectively. 

\begin{figure}[H]
\centering 
\includegraphics[scale=0.35]{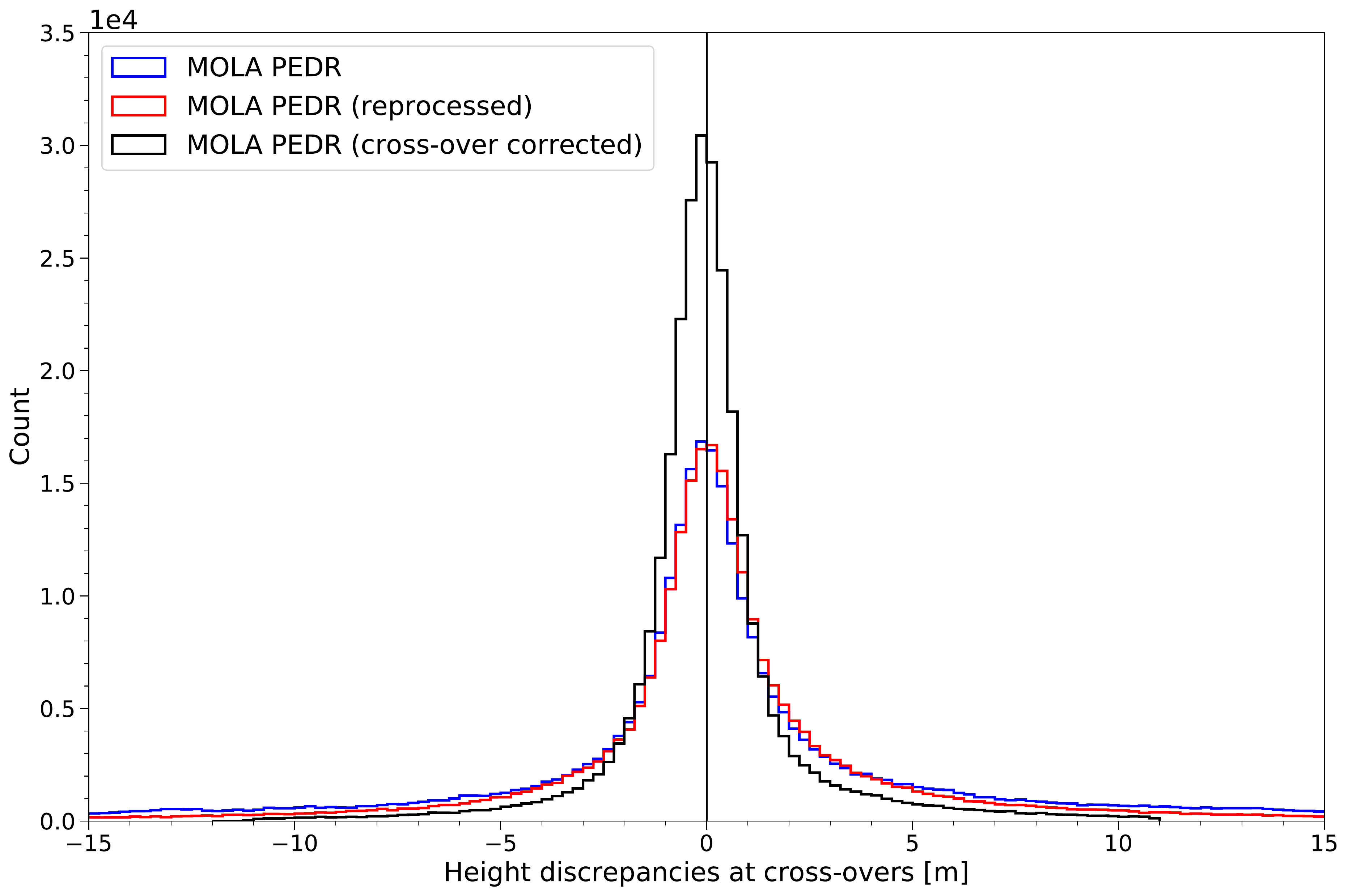}
\caption{Histogram of height discrepancies at cross-overs at [60$\degree$S,$\,$60$\degree$N,$\,$180$\degree$E,$\,$220$\degree$E] for MOLA PEDR (blue), reprocessed MOLA PEDR (red), and cross-over corrected MOLA PEDR (black).}
\label{fig:height_misfits_crossover}
\end{figure}

\quad The histograms of these cross-over height discrepancies for these two datasets are shown for comparison in Fig.~\ref{fig:height_misfits_crossover}. While the median of the height discrepancies at cross-overs stays largely unchanged (0.07~m to 0.14~m), the RMS height discrepancy has decreased from 7.1~m to 5.1~m, indicating some discrepancies in MOLA PEDR have been reduced to some extent by the incorporations of recent MGS orbit model and Mars rotational model. In addition, we also compare the aforementioned results to that of the cross-over corrected MOLA PEDR which has incorporated the corrections from the global cross-over analysis \citep{neumann2001cross-over}. The corresponding histogram is much more concentrated with a median of -0.06~m and a RMS of 2.2~m from 287,584 cross-overs, showcasing the much improved internal consistency of the dataset than the reprocessed MOLA PEDR.

\section{Internal consistency of the reprocessed MOLA PEDR after self-registration}
\label{sec:cross-over_misfits_self_registered}
\quad To investigate the internal consistency of the reprocessed MOLA PEDR after self-registration, we compare the height discrepancies at cross-overs for the reprocessed MOLA PEDR before and after the self-registration in the polar region of [78$\degree$S,$\,$87$\degree$S,$\,$160$\degree$E,$\,$220$\degree$E]. We limit the profiles acquired within March 1999 to November 1999 to reduce the computation load. To remove outliers, the initial height discrepancies are edited through a 5-$\sigma$ iterative filtering process. Consequently, 1.8$\%$ and 1.9$\%$ of these height discrepancy measurements are excluded for the reprocessed MOLA PEDR (392,688 remain) and the reprocessed MOLA PEDR after self-registration (376,002 remain), respectively. The much less number of remaining cross-overs for the the latter is due to an extra of 28 profiles being excluded in the self-registration process (see Sec.~\ref{sec:self_registration_MOLA}).

\begin{figure}[H]
\centering 
\includegraphics[scale=0.35]{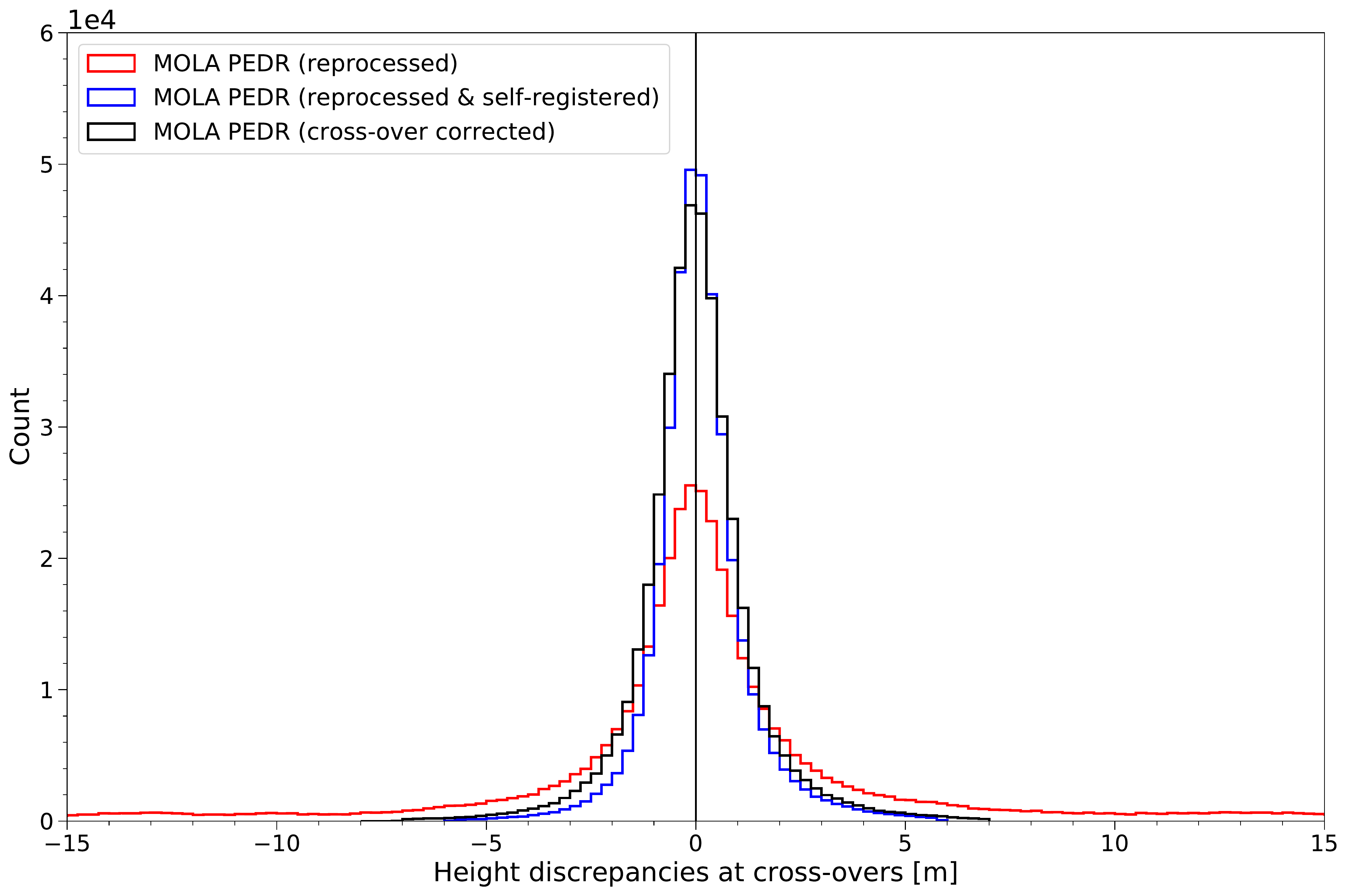}
\caption{Histogram of height discrepancies at cross-overs at [78$\degree$S,$\,$87$\degree$S,$\,$160$\degree$E,$\,$220$\degree$E] for reprocessed MOLA PEDR (red), reprocessed and self-registered MOLA PEDR (blue), and cross-over corrected MOLA PEDR (black).}
\label{fig:height_misfits_crossover_self_registered}
\end{figure}

\quad The histograms of the cross-over height discrepancies for these two datasets are compared in Fig.~\ref{fig:height_misfits_crossover_self_registered}. While the median of the height discrepancies at cross-overs stays close to 0 (0.02~m to 0.03~m), the RMS height discrepancy has substantially reduced from 5.5~m to 1.1~m, which indicates the effectiveness of the self-registration procedure. One source of the remaining internal inconsistency is the assumption of constant 3D offsets for each of the laser profile in the self-registration process with which spatially and temporally inhomogenerous seasonal height variations of the seasonal polar cap can interfere. Additionally, as in Sec. \ref{sec:cross-over_misfits_reprocessed}, we also include into the comparison of the MOLA PEDR post-corrected by the global cross-over analysis \citep{neumann2001cross-over}. The histogram of the 427,745 cross-over height residuals is represented by the black curve in Figure \ref{fig:height_misfits_crossover_self_registered}, with a median of -0.02~m and a RMS of 1.4~m. Thus, the internal consistency of the self-registered MOLA profiles, which are gridded to make the self-registered MOLA reference DTM, is close to or even better than the cross-over corrected MOLA PEDR.

\section{Height change time series using HRSC DTMs}
\label{sec:height_change_HRSC_appex}

\quad The High-Resolution Stereo Camera (HRSC) onboard Mars Express is a camera developed at the German Aerospace Center (DLR) Institute of Planetary Research and is the first stereo camera employed in a planetary mission. It comprises nine channels/looks that can collect multi-angular images of the Martian surface, allowing for along-orbit stereographic mapping \citep{gwinner2016hrsc}. Recently, \cite{putri2019new} has produced 33 DTMs with a spatial resolution of 50~m/pixel based on HRSC stereo imagery. These represent the first high-resolution DTM set that covers almost the entire Martian South Pole between 78$\degree$S and 90$\degree$S. These DTMs show RMS height residual of $\sim$9.5~m compared to MOLA PEDR footprints and $\sim$7.3~m to the MOLA Mission Experiment Gridded Data Records (MEGDRs) with resolution of 512 pixels/degree \citep{putri2019new}. Here, we adopt these DTMs from optical images instead of the self-registered MOLA reference DTM as the static mean surface in "F$\_$LC"-method and the 2$^{\rm{nd}}$ Regional Pseudo Cross-over Adjustment (RPCA) in the bi-RPCA post-correction procedure, and try to assess the feasibility and performance of using these HRSC DTMs in deriving the seasonal height variations.

\begin{figure}[H]
\centering 
\includegraphics[scale=0.95]{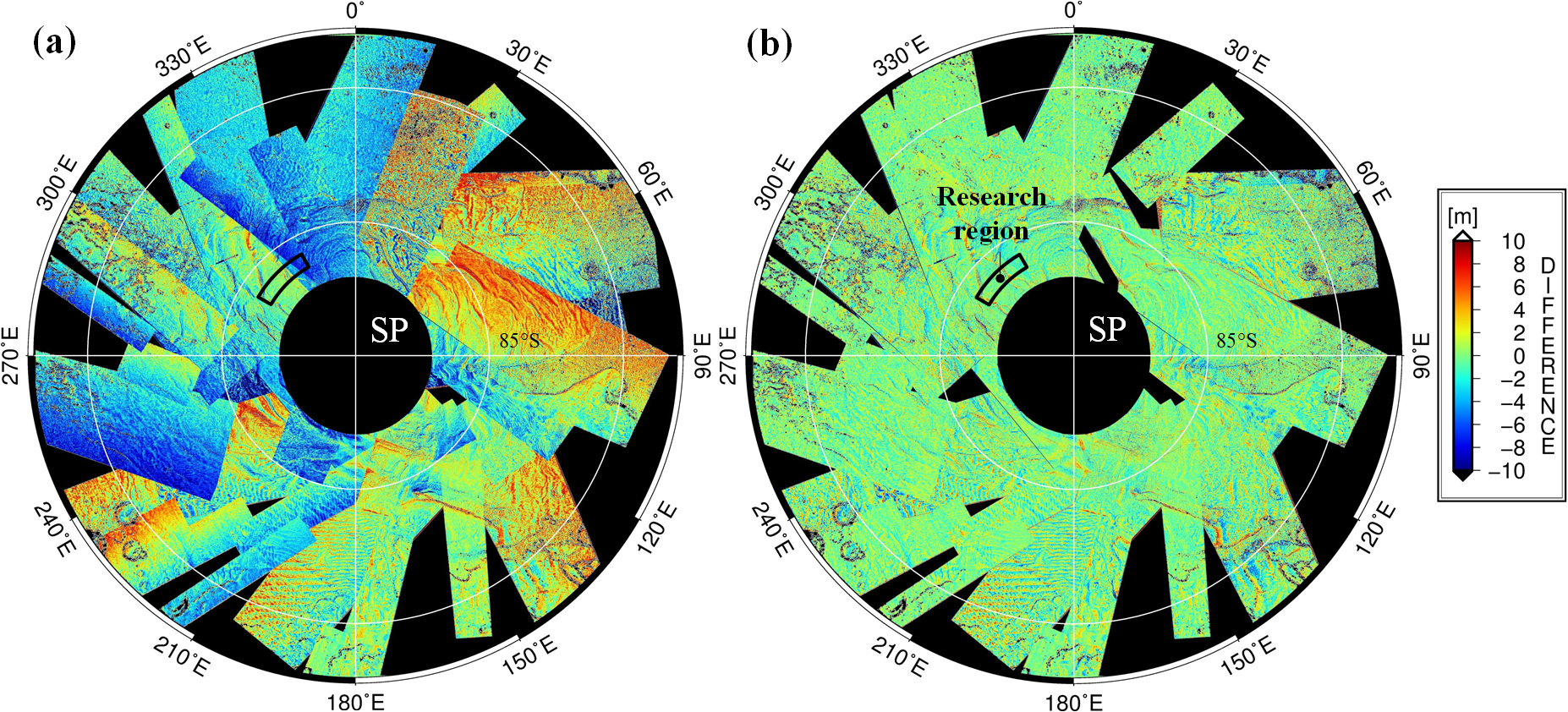}
\caption{(a) Height differences between HRSC DTM mosaic and the self-registered MOLA profiles before the alignment. Relative tilts of magnitudes up to $\sim$10~m are ubiquitous. (b) Height differences between HRSC DTM mosaic and the self-registered MOLA profiles after the alignment. Differences are homogeneous with magnitudes typically less than 2~m except in some rough regions at the periphery from 78$\degree$S to 82$\degree$S. The linear patterns visible between 180$\degree$E and 210$\degree$E are likely to be errors in the homologous image point matching processes when generating the HRSC DTMs. The fan-shaped regions outlined on both plots are the research region for this study. The projection is the same as in Fig.~\ref{fig:MOLA_slope_map}. This figure is plotted using the Generic Mapping Tool (GMT) software \citep{Wessel2013GMT}.}
\label{fig:MOLA_height_dif}
\end{figure}

\begin{figure}[H]
\centering 
\includegraphics[scale=0.32]{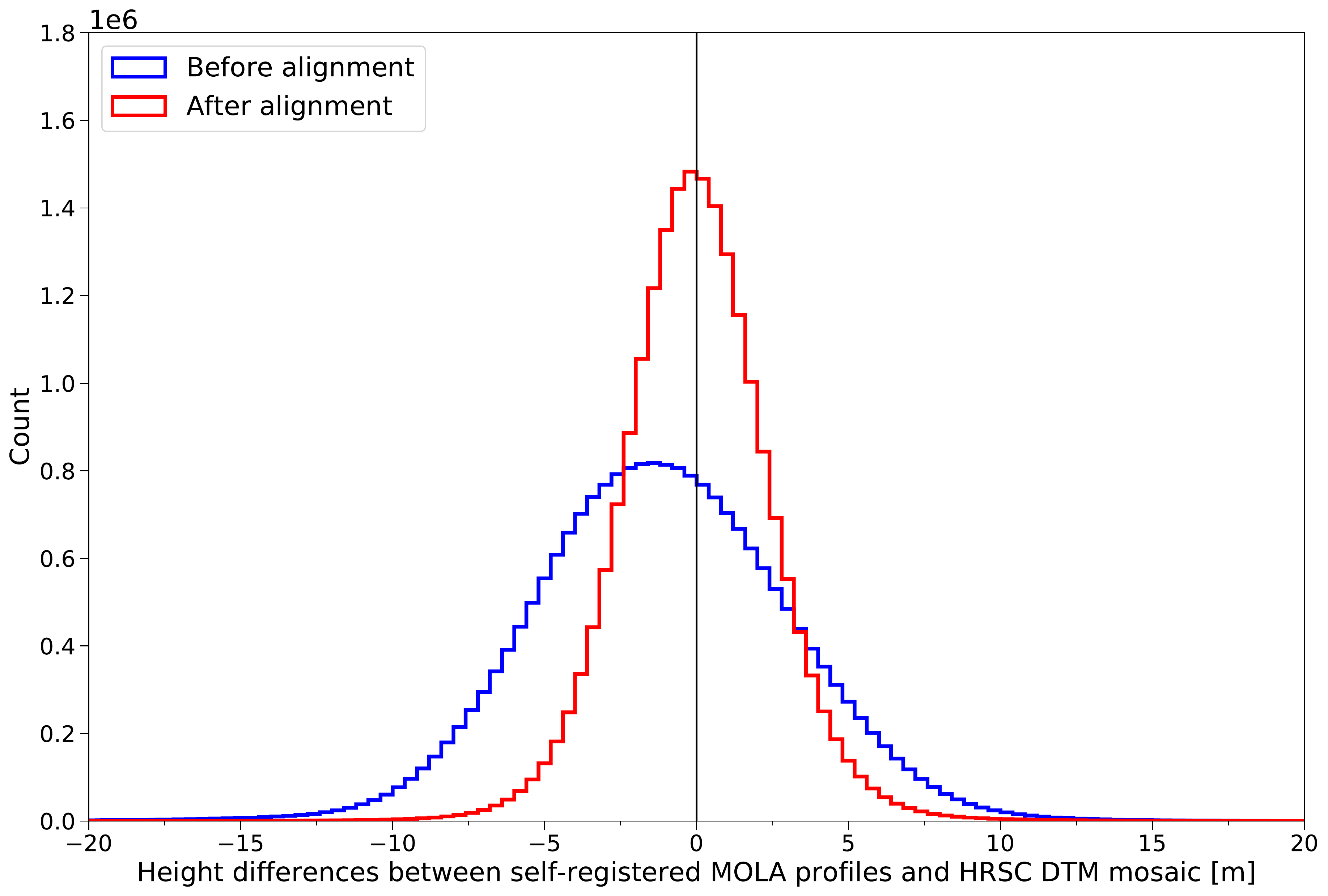}
\caption{Histogram of the height differences between HRSC DTM mosaic and the self-registered MOLA profiles before and after the alignment. The RMS height difference has decreased from 4.4~m to 2.5~m.}
\label{fig:MOLA_height_dif_hist}
\end{figure}

\quad These HRSC DTMs were created without geodetic control and are prone to overall tilts and offsets \citep{putri2019new}. Thus, these DTMs have to be tied to the geodetic reference represented by the self-registered MOLA profiles in Sec.~\ref{sec:self_registration_MOLA} before being utilized. If the misalignment was left unattended, the generated height change time series can be significantly biased and unbalanced, e.g., continuously increasing or deceasing with unreasonable rates. We carry out the alignment of the individual HRSC DTMs to the self-registered MOLA profiles using the \texttt{pc$\_$align} tool of the Ames Stereo Pipeline \citep[ASP;][]{beyer2018asp}. This program uses an Iterative Closest Point (ICP) matching algorithm to derive the transformation parameters between a point cloud and a DTM. For robustness, \texttt{pc$\_$align} keeps 75$\%$ of the points with the smallest errors for the estimation of a combination of rotation and translation parameters. In the effort to create an aligned HRSC DTM mosaic, we deliberately exclude two DTMs named \texttt{h4837$\_$0000$\_$dt4$\_$ucl} and \texttt{h4917$\_$0009$\_$dt4$\_$ucl} due to their much worse quality compared to the others. The top-to-bottom order of the remaining DTMs during mosaicing remains unchanged as to \cite{putri2019new} except that a DTM named \texttt{h2287$\_$0000$\_$dt4$\_$ucl} is lifted to the top layer due to its best quality at its coverage. Fig.~\ref{fig:MOLA_height_dif} and Fig.~\ref{fig:MOLA_height_dif_hist} show the comparison of the height differences between the HRSC DTM mosaic and the self-registered MOLA profiles before and after the alignment, respectively. Tilts and shifts of several meters are ubiquitous before the alignment, while these become much less significant after the alignment. The RMS height difference has decreased from 4.4~m to 2.5~m, an improvement of roughly 40$\%$. In addition, a -1.2~m median height offset between the two data types has been compensated via the alignment process.

\label{sec:aligning}

\begin{figure}[H]
\centering
\includegraphics[scale=1]{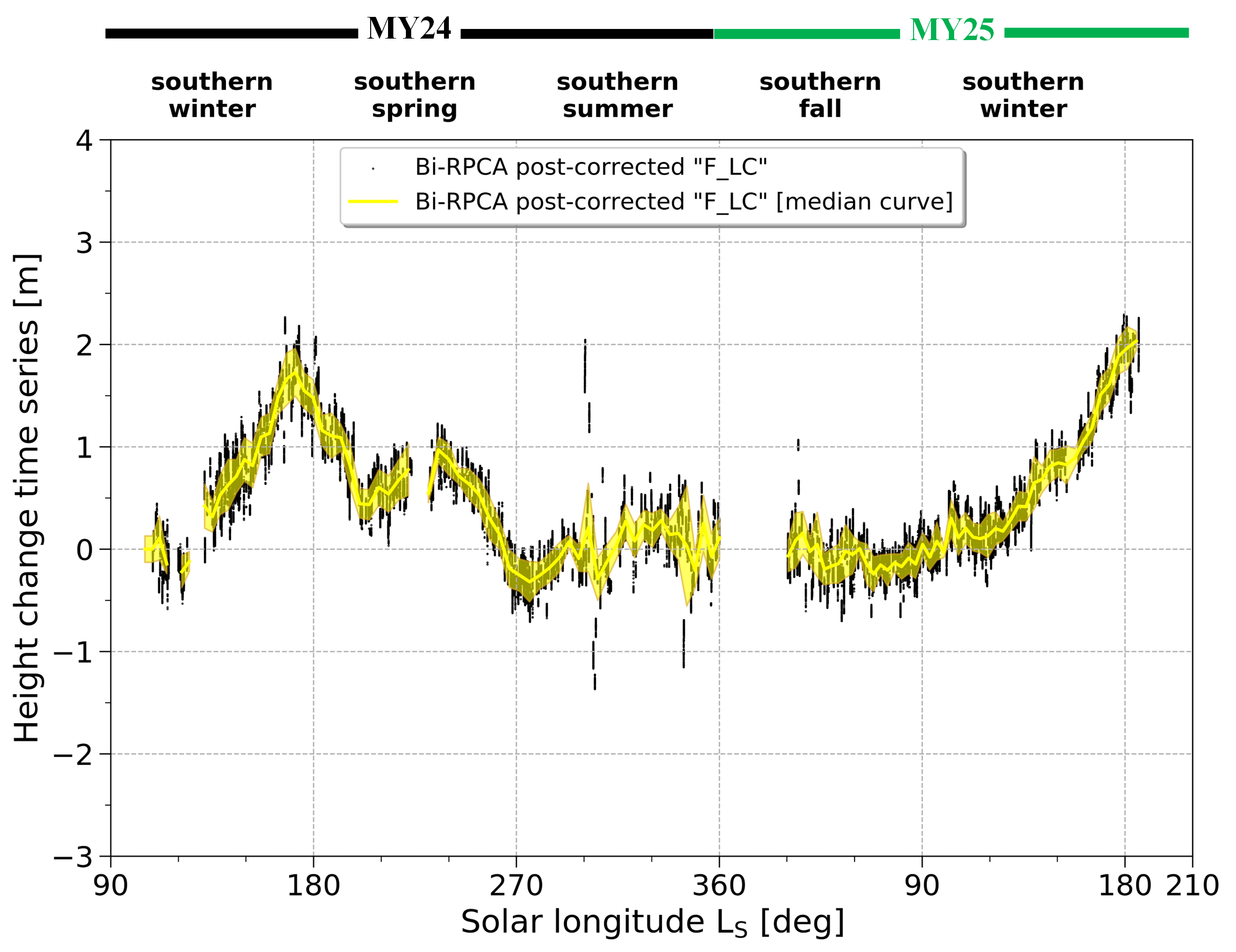}
\caption{Temporal height variation of the research region derived from "F$\_$LC"-method post-corrected by the bi-RPCA procedure using the aligned HRSC DTM mosaic as the reference surface.}
\label{fig:height_change_bin_region_HRSC}
\end{figure}

\quad Then, we apply the aligned HRSC DTM mosaic as the baseline surface to "F$\_$LC"-method complemented with the bi-RPCA procedure in an attempt to map the seasonal height variation at the same research region. The optimal regularization strength in the 2$^{\rm{nd}}$ RPCA is set to 1.5 according to the ridge trace analysis. The derived height change time series is shown in Fig.~\ref{fig:height_change_bin_region_HRSC}. A generally consistent trend has been captured as that using the self-registered MOLA reference DTM (compare to bottom panel of Fig.~\ref{fig:height_change_bin_region_F}), demonstrating the feasibility of using the DTMs from optical stereo images in mapping the seasonal height variations. However, the mean precision as measured by $\scli{\mathrm{\small MAD}}{s}$ is 17~cm which more than triples compared to 4.9~cm in the case involving the self-registered MOLA reference DTM.


\end{document}